\documentclass[14pt,preprint]{aastex}

\newcommand{\um}{$\mu$m}
\newcommand{\brgamma}{Br$\gamma$}
\newcommand{\kms}{km\thinspace s$^{-1}$}

\def\degr{\hbox{$^\circ$}}
\def\arcmin{\hbox{$^\prime$}}
\def\arcsec{\hbox{$^{\prime\prime}$}}
\def\utw{\smash{\rlap{\lower5pt\hbox{$\sim$}}}}
\def\udtw{\smash{\rlap{\lower6pt\hbox{$\approx$}}}}

\def\Lsun{\hbox{\it L$_\odot$}}
\def\Lstar{\hbox{\it L$_*$}}
\def\Lx{\hbox{\it L$_X$}}

\def\Msun{\hbox{\it M$_\odot$}}

\def\Teff{\hbox{\it T$_{\rm eff}$}}

\def\J{\hbox{\it J}}
\def\H{\hbox{\it H}}
\def\K{\hbox{\it K}}

\def\Mk{\hbox{\it M$_{\rm K}$}}

\newcommand{\Ks}{{\it K$_{\rm s}$}}

\newcommand{\Al}{{\it A$_\lambda$}}
\newcommand{\Aks}{{\it A$_{\rm K_{\rm s}}$}}

\def\BCK{\hbox{\it BC$_{\rm K}$}}
\def\BCV{\hbox{\it BC$_{\rm V}$}}
\def\simgr{\mathrel{\hbox{\rlap{\hbox{\lower4pt\hbox{$\sim$}}}\hbox{$>$}}}}

\shorttitle{Massive stars in direction of the W33 HII region.}
\shortauthors{Messineo et al.}

\begin{document}


\title{Massive stars in the  Cl 1813-178 Cluster.  
An episode  of massive star formation in the W33 complex.}


\author{Maria~Messineo\altaffilmark{1,2}, 
	Ben~Davies\altaffilmark{3,5,10},
        Donald~F.~Figer\altaffilmark{3},
        R.P.~Kudritzki\altaffilmark{7,9},
	Elena Valenti \altaffilmark{4},
	Christine Trombley\altaffilmark{3},        
	F.\ Najarro \altaffilmark{8},
        R. Michael Rich\altaffilmark{6} 
}
 
\email{messineo@mpifr-bonn.mpg.de}

\altaffiltext{1}{European Space Agency (ESA). The Astrophysics and Fundamental Physics Missions Division, Research and
Scientific Suppport Department, Directorate of Science and Robotic Exploration,
ESTEC, Postbus 299, 2200 AG Noordwijk, the Netherlands.}
\altaffiltext{2}{ Max-Planck-Institut fuer Radioastronomie, Auf dem Huegel 69, D-53121 Bonn. }

\altaffiltext{3}{Center for Detectors, Rochester Institute of Technology, 74 Lomb Memorial Drive, 
   Rochester, NY 14623, USA}

\altaffiltext{4}{European Southern Observatory, 
Karl Schwarzschild-Strasse 2, D-85748 Garching bei Munchen, Germany}

\altaffiltext{5}{School of Physics \& Astronomy, University of Leeds, Woodhouse Lane, Leeds
LS2 9JT, UK.}

\altaffiltext{6}{Physics and Astronomy Building, 430 Portola Plaza, Box 951547, Department of Physics and Astronomy, University of California, Los Angeles, CA 90095-1547.}

\altaffiltext{7}{Institute for Astronomy, University of Hawaii, 2680 
Woodlawn Drive, Honolulu, HI 96822}

\altaffiltext{8}{Centro de Astrobiolog\'{\i}a (CSIC-INTA),
Ctra. de Torrej\'on a Ajalvir km4, 28850, Torrej\'onde Ardoz, Madrid, Spain}

\altaffiltext{9}{
Max-Planck-Institute for Astrophysics, Karl-Schwarzschild-Str. 1, 85748 Garching, Germany}

\altaffiltext{10}{
Institute of Astronomy, University of Cambridge, Madingley Road, Cambridge, CB3 0HA, UK
}


%

\begin{abstract}
Young massive ($M >10^4$ \Msun) stellar clusters are a good laboratory to study
the  evolution of massive stars. Only a dozen of such clusters are known in the Galaxy.
Here we report about a new young massive stellar cluster in the Milky Way.
Near-infrared medium-resolution spectroscopy with UIST on the UKIRT telescope and NIRSPEC on  
the Keck telescope, and X-ray observations with the Chandra and XMM satellites, of the Cl 1813-178 
cluster confirm a large number of massive stars.
We detected 1 red supergiant, 2 Wolf-Rayet stars, 1 candidate luminous blue variable, 
2 OIf,  and 19 OB stars. Among the latter, twelve are likely supergiants, four  giants, and 
the faintest three dwarf stars. We detected  post-main sequence stars with masses 
between 25 and 100 \Msun.  A population with age  of 4-4.5 Myr and a mass of $\sim 10000$ \Msun\
can reproduce such a mixture of massive evolved stars.  
This massive stellar cluster is the first  detection of a cluster in the W33 complex. 
Six  supernova remnants and several other candidate clusters are found in the 
direction of  the same complex. 

\end{abstract}





\section{Introduction}    

An understanding of the mechanisms of  formation, evolution, and end state of massive stars is  fundamental
for the studies of galaxies at all redshifts. Massive stars contribute to the chemical enrichment of
the interstellar medium with their strong winds and by exploding as  supernovas. Massive stars are the most
luminous stars, can easily be detected in external galaxies, and provide  distance estimates. They
are the sources of the most energetic phenomena in the Universe,   gamma ray bursts
\citep[e.g.][]{woosley06}.


The availability of large surveys of the Galactic plane at  radio and infrared   wavelengths  opens a
golden epoch for studying  the formation, evolution, and environments of massive stars.   More than
1500 new candidate stellar clusters have been discovered, and  among them several young  clusters rich
in massive stars may be hidden \citep{messineo09a}. 

In \citet{messineo08} (hereafter referred to as PaperI),  we presented the serendipitous discovery of a young massive cluster, Cl 1813-178,
in the Galactic disk  at l=12\degr, with a spectroscopically identified population of massive stars, 
a red supergiant star (RSG), two blue supergiants (BSG), and one Wolf-Rayet (WR) star.  Here, we
present a follow-up study  of the  cluster. Near-infrared photometry and  spectroscopy,  and
X-ray observations,  reveal a large number of massive stars.

The cluster is located  in the  W33 complex, and is associated with two supernova remnants (SNR), 
SNR G12.72$-$0.00 and G12.82$-$0.02, and the highly magnetized pulsar associated with the TeV 
$\gamma$-ray source HESS J1813$-$178. Interestingly, the W33 complex  appears to contain  several
other candidate stellar clusters, and several  SNRs.  Clusters  do form  in large complexes
\citep[e.g.][]{beuther07}, and  their spatial distribution  varies from cloud to cloud,   indicating
that several  external and internal triggers can be at work \citep[e.g.][]{clark09}.  W33 is an ideal
laboratory to investigate  various issues about massive stars and multi-seeded star formation, and to
verify the  presence of triggered sequential star formation, which is  suggested by the presence of 
SNRs. The association of the stellar cluster  with two   SNRs  can shed light on the initial masses
of  the supernova progenitors, and  on the fate of massive stars. 

We describe the observations and data reduction in Sect.\ 2, and  we analyze the
spectra and the cluster color-magnitude diagram (CMD) in Sect.\ 3. 
A discussion on the massive members and spectrophotometric distances is presented in Sects.\ 4 and 5.  
Cluster age and mass are
derived in Sects.\ 6 and 7. An overview of candidate clusters in the direction of the 
W33 complex is given in Sect.\ 8. Finally,  our findings are summarized  in Sect.\ 9.

\section{Observations and data reduction}

\subsection{IR photometry}
Photometric measurements from the near-infrared Two Micron All Sky Survey (2MASS) catalog
\citep{skrutskie06},  the Galactic Legacy Infrared Mid-Plane Survey Extraordinaire (GLIMPSE) 
catalog \citep{benjamin03}, as well as from the Naval Observatory Merged Astrometric Dataset 
(NOMAD) \citep{zacharias04} were used.

Images of the central cluster region with the  $J$, $H$ and \Ks\ filters were obtained during two
nights of observation, 2008 June 23-24\footnote{Based on data taken within the observing program
081.D-0371.}, using the near\--IR camera SofI mounted on the ESO NTT \citep{moorwood98}. We used
SofI in large\--field mode with a pixel size of 0.288\arcsec, and a total field of view of
4.9'$\times$4.9'.  For each filter, we performed a random dithering pattern of 14 images, and
reached  a total exposure time of 840s, 1176s, and 1680s in $J$, $H$ and \Ks, respectively. Each
image is a combination of 50, 70, and 100 exposures, each one 1.2~s long, in $J$, $H$, and \Ks\
bands, respectively. 

Data reduction was performed using standard IRAF routine.  For each filter, we obtained a sky image 
by  median-combining  the dithered frames, and we subtracted the sky  image from each frame,  
Flat fielding was performed by using the {\it SpecialDomeFlat} template, which applies the 
appropriate illumination correction, as described in the SofI User Manual. Finally, the 
dithered frames were averaged into a
single image. Standard crowded-field photometry, including point-spread function modeling, was
carried out on each image using DAOPHOT~II/ALLSTAR (Stetson 1987). The internal photometric accuracy
was estimated from the {\it rms} frame\--to\--frame scatter of multiple star measurements
(0.03~mag$<{\sigma}_J$$ {\sim}{\sigma}_H {\sim}{\sigma}_{K_s}<$0.06~mag for
8$<J{\sim}H{\sim}K_s<$18).  The instrumental magnitudes were  converted into the 2MASS photometric
system\footnote{An overall uncertainty of $\pm$0.05~mag in the zero\--point calibration in all the
three bands has been estimated}. The 2MASS catalog was used as an astrometric reference 
frame\footnote{The astrometric procedure provided {\it rms} residuals of $\approx$0.2\arcsec\ in both
the right ascension and declination.}. For stars that were saturated on the SofI images, magnitudes were
taken from the 2MASS catalog.

\begin{figure*}[!]
\begin{centering}
\resizebox{0.9\hsize}{!}{\includegraphics[angle=0]{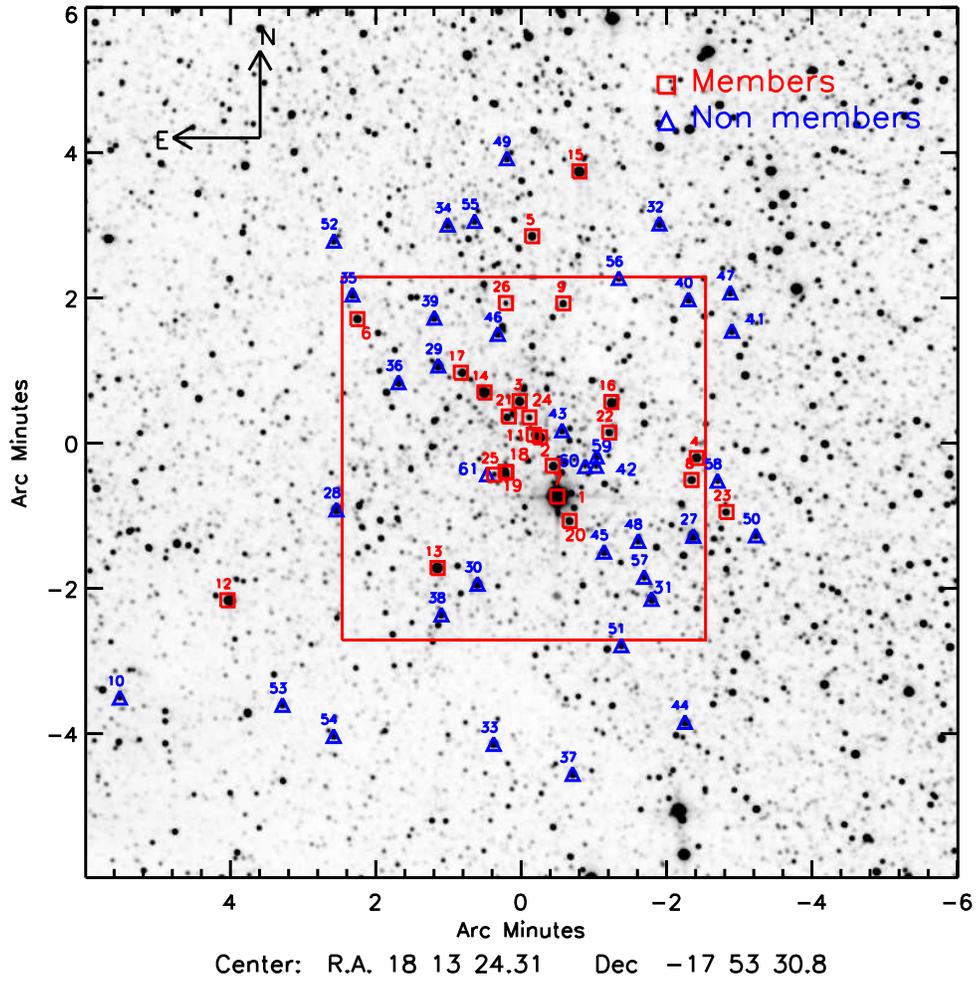}}
\end{centering}
\caption{\label{fig.map} 2MASS \Ks-band image of the cluster. The big box shows the 
location of the field observed with SofI. Stars that were
spectroscopically observed and are likely cluster members are marked with small boxes. 
Non-member stars are marked with triangles.}
\end{figure*}

\subsection{Near\--IR spectroscopy}
Spectroscopic observations of the  brightest  stars   were carried
out  with NIRSPEC at the Keck Observatory\footnote{Data
presented herein were obtained at the W.M. Keck Observatory, which is operated as a
scientific partnership among the California Institute of Technology, the University of
California and the National Aeronautics and Space Administration.  The Observatory was
made possible by the generous financial support of the W.M. Keck Foundation.}  under program  H243NS (PI: Kudritzki) on
2008 July 24. The \K-filter and a 42\arcsec $\times$ 0.570\arcsec\ slit  were
used, covering from 2.02 to 2.45 \um\ with a  resolution of R=1700. For each target,
two nodded exposures of 10s each were taken. We observed a total of 23 stars with
NIRSPEC  (see Tables \ref{table.members} and \ref{table.nonmembers}).

Additional spectroscopic data were taken with the UKIRT 1-5 micron  Imager
Spectrometer (UIST) at the  UKIRT Observatory \footnote{The United Kingdom Infrared Telescope (UKIRT) is
operated by the Joint Astronomy Centre on behalf of the Particle Physics and Astronomy
Research Council.}. We used the short-K grism in combination with a 120\arcsec $\times$
0.12\arcsec\ slit, covering  from 2.00 to 2.26 \um\ at a resolution  R=1800. We also used
the long-K  grism to cover the spectral range  from 2.204 to 2.513 \um\ at a
resolution of R=1900.  Typically, each target was observed with the long-K grism. When
CO bands were not detected, a second spectrum was taken with   the short-K grism.
Integration times  varied from 30 to 60 s per exposure, and the number of exposures
varied from 8 to 20 s. The 39 stars observed with UKIRT, including a few chance
detections,  are listed in Tables \ref{table.members} and \ref{table.nonmembers}.

Pairs of frames with  nodded positions were subtracted and flat-fielded. The stellar
traces were straightened  using a two dimensional de-warping procedure. We wavelength
calibrated the spectra with arc lines. We corrected the spectra for atmospheric absorption and
instrumental response by dividing the observed spectra by the spectrum of a
reference star. Reference stars were  of B type (from B2 to B9). The \brgamma\
and HeI lines of the telluric spectrum were eliminated with linear interpolation.

The signal to noise of the spectra varied from 40 to 150. 

\subsection{X--ray data}
The cluster is located in the vicinity of the HESS J1813-178 pulsar wind nebula \citep{helfand07}.
Several  X-ray observations of this region have been performed in order to identify the pulsar 
associated with the  HESS source.  \citet{helfand07}
presented a catalog of 75 point sources  detected with the Chandra satellite. We
cross-correlated the Chandra catalog with  the 2MASS point source catalog, and identified  a
total of 44 matches (PaperI). 
Using the XMM satellite, \citet{funk07}  detected seven X-ray  sources 
in the surrounding area of HESSJ1813-178.
 
\section{Analysis}

\subsection{Spectral classification}
Spectral classification was performed by comparing the spectra with spectral atlases
\citep[e.g.][]{hanson96,hanson05, figer97, martins07, kleinmann86, wallace96}.

\begin{figure}[!]
\begin{centering}
\resizebox{0.8\hsize}{!}{\includegraphics[angle=0]{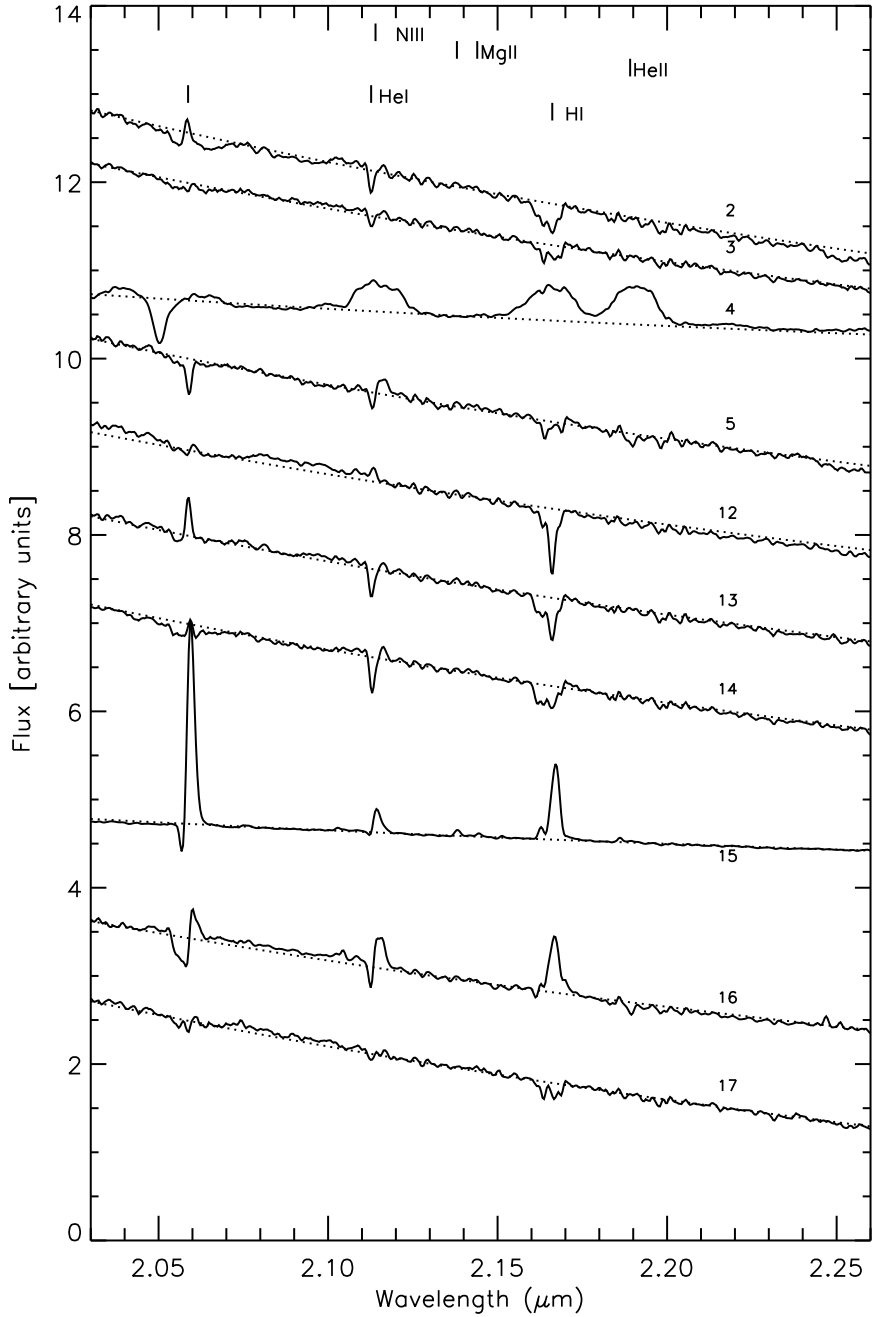}}
\end{centering}
\caption{\label{fig.keckearly} K-band spectra of early-type stars obtained with NIRSPEC
on the Keck telescope. 
The full line shows the spectrum after correction for atmospheric 
absorption and instrumental response, and interstellar extinction. 
Extinction correction was performed with the values listed in Table \ref{table.properties}, 
and by assuming a power law with an index of $-$1.9 \citep{messineo05}.
The dashed line shows a black body with a temperature equal to the stellar 
effective temperature.
Each spectrum was normalized at 2.12 \um, and arbitrarily shifted for clarity.
The locations of detected spectral lines are indicated with marks and labels. }
\end{figure}

\begin{figure}[!]
\begin{centering}
\resizebox{0.8\hsize}{!}{\includegraphics[angle=0]{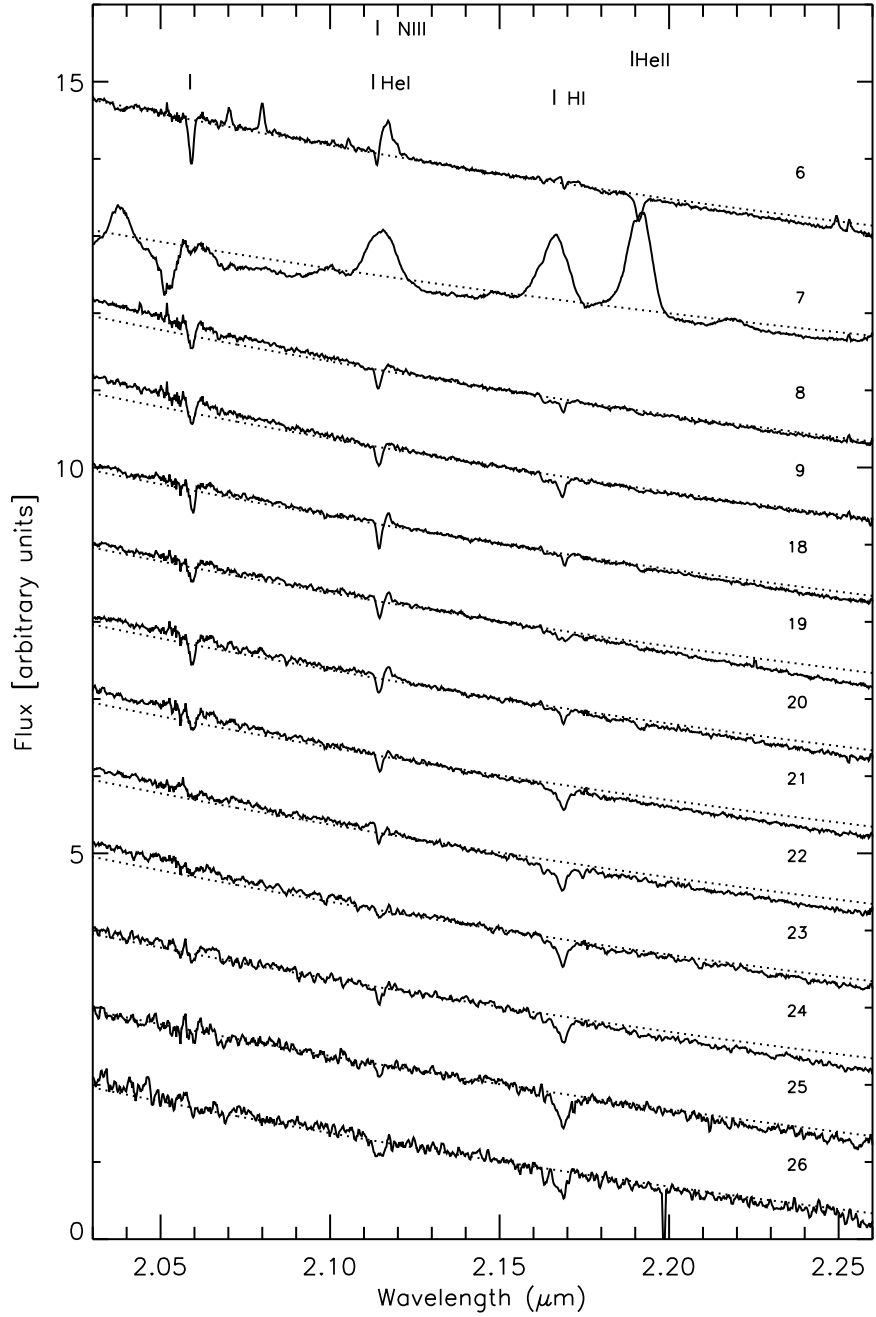}}
\end{centering}
\caption{\label{fig.ukirtearly} K-band spectra of early-type stars obtained with UIST 
on the UKIRT telescope. Curves and labels are as in Fig.\ \ref{fig.keckearly}}.
\end{figure}

We observed a total of 60 stars, and detected 24 early-type stars and 36 late-type stars.
In addition, we considered the previous observations reported in PaperI,  
which added an extra early-type star (\#11) to our new sample.
The observed stars were divided into candidate cluster members and non members, 
and are listed in Tables \ref{table.members}, and  \ref{table.nonmembers}.
We first listed the sample of stars given in the  Table 1 of PaperI, 
using the same identification numbers, then we appended the new targets
ordered in \Ks\ magnitude. The spectra of the newly detected early-type stars are 
shown in Figs. \ref{fig.keckearly}, and \ref{fig.ukirtearly}.

In the following, we describe  the spectra by grouping the stars into late-types
and early-types. Early types  divided in four subgroups (Wolf-Rayet stars, 
candidate luminous blue variables, O-type, and B-type stars). 

\subsubsection{Red Supergiants}
In the \K-band, the spectra of late-type stars  have CO band-heads at 2.29 \um.  
By measuring the equivalent widths of the CO bands,
it is possible to  distinguish between giants and supergiants (see the Appendix).

Among the detected late-type stars, four stars, \#1, \#32, \#38, and  \#39, are found to 
have equivalent widths of the CO bands, EW(CO), typical of RSGs.

Star \#1 is a K2-K5I cluster member, as discussed  
in PaperI on the basis of high-resolution data.

Star \#32, \#38, and  \#39 have EW(CO)s  typical of RSGs with spectral types
M2.5, M3.5 and M1, respectively. Their photometric properties
indicate that they are unrelated to the stellar cluster.
We  further discuss these stars in the following sections.

\subsubsection{Wolf-Rayet stars}

\begin{figure}[!]
\begin{centering}
\resizebox{0.8\hsize}{!}{\includegraphics[angle=0]{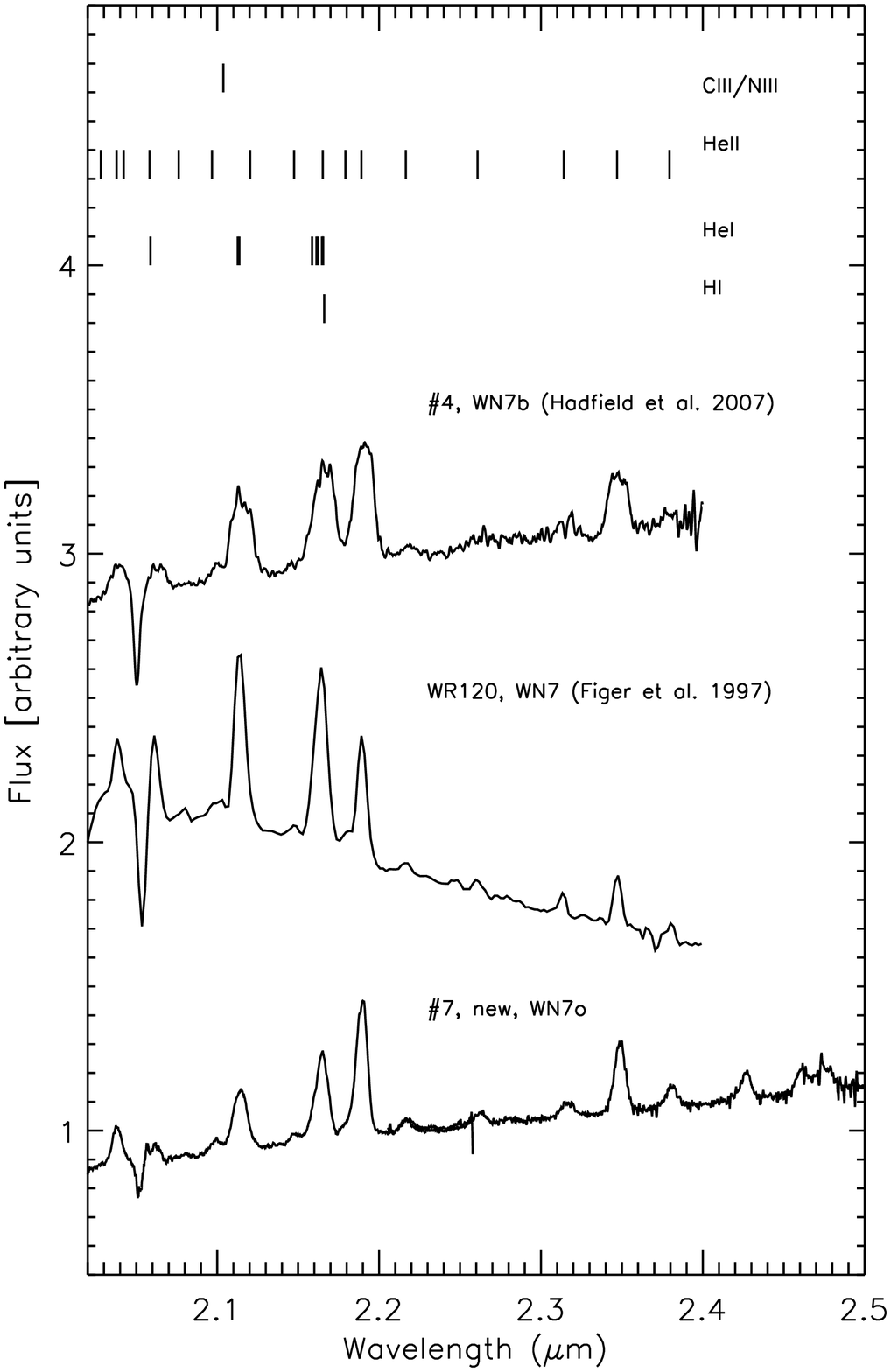}}
\end{centering}
\caption{\label{fig.wr} Spectra of star \#4, which is an already known 
WR star \citep{hadfield07}, and star  \#7, which is a newly detected WR star.
For comparison, a spectrum of a WN7 star, WR120, is presented from the atlas of
\citet{figer97}. The rest wavelengths of  HeII, HeI, HI, and CIII/NIII lines are also marked.}
\end{figure}

Star \#4  is a known WR star of type WN7 \citep[number \#8
in][]{hadfield07}. Star \#7  has similar  emission lines at 2.114 \um,   2.166 \um, and
2.189 \um.  By comparing it with the spectral atlas of \citet{figer97}, 
star \#7 appears to be another WN7 star (see Fig.\ \ref{fig.wr}).

\subsubsection{Candidate Luminous Blue Variables}

The spectrum of star \#15  is dominated by He I lines at 2.058 \um\  and 2.112 \um, and   \brgamma\ line
in emission. Faint  MgII lines are detected at 2.138 \um\ and 2.144 \um. The type of
detected lines and the  line profiles indicate that star \#15 is a P-Cygni B supergiant.
Its  spectrum is similar to that of the candidate luminous blue variable (cLBV) star 
in the W51 region \citep[][]{clark09}.  More information on the stellar photometric variability 
is needed to firmly confirm that it is another LBV star.

\subsubsection{O-type stars}

Spectra of O and B stars have hydrogen lines, helium lines, and other atomic lines 
(e.g. CIV, NIII, CIII, MgII).  Hydrogen and helium lines are usually seen in absorption;
they may be in emission in supergiants. The \brgamma\ line is the only hydrogen line
detectable in K-band. The HeI transitions at 2.058\um\ and 2.112 \um\ are usually detected
in late-O and early-B stars. The HeII line at 2.189 \um\ appears 
only in O-type stars. With our resolution, the HeII line at 2.189\um\  may be present   
down to O9.5I or O9V.

The spectrum of star \#6 has a strong HeI line at 2.058 \um\ in absorption,  CIV lines at 2.069 \um\ and 2.079 \um\
in  emission,  strong emission line at 2.11 \um\ (probably due to HeI/NIII/CIII), and a HeII absorption line
at 2.189 \um. NIII at 2.25\um\ is seen in emission.
There is  an indication for a faint CIV line at 2.083 \um, and a NIII line at 2.10
\um.  The simultaneous presence of HeI, HeII, and CIV lines is typical of an O-type star. The strengths of the
HeI and HeII absorption lines, as well as the detection of  NIII at 2.25\um, indicate a O6-O7f+.

The spectrum of star \#16 has \brgamma\ line in emission,  a HeI line at 2.058 \um\ with P-Cygni profile,
and a HeI/NIII line at  2.112 \um. A HeII line at 2.189 \um\ is 
seen in absorption. A faint line is seen in emission at 2.105 \um. Only an indication for a CIV 
line at 2.08 \um\ is visible. The spectrum resembles that of  a O8-O9If  star, similar to
HD151804 \citep{morris96}. 

Stars  \#18, \#19, and \#20 have HeI lines at 2.058 and 2.112 \um\ in absorption,  a NIII line  
at 2.115 \um\ in emission, and \brgamma\ line in absorption. A HeII line at 2.189 \um\ clearly appears  
in absorption.   The absence of CIV lines,  and the presence of NIII 
and HeII lines, suggest a late O-type  (O7-09). 

The spectra of stars \#5, \#8, and \#9    have  HeI lines at 2.058 \um, and  2.112 \um\ in 
absorption, and \brgamma\ line in absorption. There is an indication for
a NIII line at 2.115 in emission, and a  HeII line at 2.189 \um\ in absorption.
Star \#5 has a clear detection  of a NIII line.  
The absence of CIV lines, and the presence (or indication) of NIII and HeII lines, suggest 
that this is a  late O-type  (O7-O9).

\subsubsection{B-type stars}

Here, we list all detected early-type stars, which do not have  HeII lines. These may 
be B-stars or O9 dwarfs. With our resolution, the HeII line at 2.189\um\  may be present   
down to O9.5I or O9V. Once the HeII line disappears, assigning a  spectral type and 
luminosity class can turn into a quite degenerate issue. This degeneracy may be 
partly broken if the absorption components of the \brgamma\ line and of the HeI lines are sufficiently filled 
in by the stellar wind contribution. Typically, a HeI line at 2.112\um\ in absorption is 
observed down to B8Ia stars, while the same line is not detected in dwarfs/giants later than B3V. Such matters are highly
dependent on the quality of the spectra.

From a comparison with the atlas of \citet{hanson96},  
{ stars \#2, \#3, \#11, \#12, \#13, \#14,  \#17, \#21, \#22, \#23, \#24,  \#25, and  \#26 appear to be likely stars
with types from B0 to B2.}

The spectra of stars \#2 and \#13  have a HeI line at 2.058 \um\  in
emission, a HeI line at 2.112 \um,  and  \brgamma\ line in absorption. 
From the relative strengths of the \brgamma\ and HeI lines \citep{hanson96},  
the presence of HeI in emission at 2.058\um, and the lack of HeII lines, these stars 
appear to be B0-B3 supergiants. 

The spectra of stars \#3,  \#22,  \#24 , \#25,  and \#26 have a HeI line at 2.112
\um\   in absorption, as well as  \brgamma\ line  in absorption.  Star \#17 
has    \brgamma\  line in absorption, and a trace of HeI
at 2.112\um.  These are likely  B0-B3 stars
or  O9-B3 stars (depending on the luminosity class).

The spectra of stars \#11, \#21, and \#23  have \brgamma\  line in absorption, HeI lines at 2.058 
and 2.112 \um\  in absorption. These spectra are typical of early  B-type stars. 

Star \#12 has   \brgamma\   line in absorption. The lack of HeI al 2.112 in absorption
suggests  that  star \#12 has a spectral type later than a B8I or B3-B4V star.

The spectrum of star \#14 has \brgamma\ in absorption,  a HeI line at 2.112 \um\  in absorption,
an indication for a HeI line in emission at 2.058 \um, and a faint   HeI line at 2.113 \um\ in emission.  
The lack of HeII at 2.189\um\ and presence of HeI lines suggest that star \#14\ is a BSG star with spectral type between  
B0 and B3.

The spectrum of star \#11 is shown in PaperI.
Stars \#2 and \#3 were also observed with NIRSPEC in April 2008 (PaperI), and we 
obtained the same spectral classification. 

\begin{deluxetable}{lrrrrrrrrrrrrll}
\rotate
\tablewidth{0pt}
\tabletypesize{\scriptsize}
\tablecaption{\label{table.members} List of cluster members spectroscopically observed.}
\tablehead{
\colhead{ID}& 
\colhead{Ra}& 
\colhead{Dec}& 
\colhead{B} & 
\colhead{V} & 
\colhead{R} &
\colhead{J} & 
\colhead{H} & 
\colhead{K$_{\rm s}$} &
\colhead{[3.6]} & 
\colhead{[4.5]} &
\colhead{[5.8]} & 
\colhead{[8.0]} & 
\colhead{ tel.} &
\colhead{Sp}    
}
\startdata
 1\tablenotemark{a}& 18 13 22.26  & -17 54 15.55  &  14.74  &  13.68  &  12.73  &   5.46  &   4.23  &   3.79  &   5.07  &   4.05  &   3.57  &  \nodata  & keck &  RSG\\
 2  & 18 13 23.26  & -17 53 26.54  &  20.08  &  15.03  &  14.50  &   8.64  &   7.72  &   7.22  &   6.94  &   6.78  &   6.63  &   6.68  &       keck &     B0-B3   \\
 3  & 18 13 24.43  & -17 52 56.75  &  20.42  &  15.74  &  14.68  &   9.20  &   8.25  &   7.79  &   7.47  &   7.38  &   7.30  &   7.30  &       keck &     B0-B3   \\
 4  & 18 13 14.19  & -17 53 43.60  &  20.48  &  17.57  &  15.20  &   9.62  &   8.60  &   7.94  &   7.34  &   6.93  &   6.70  &   6.35  &       keck &     WN7	  \\
 5  & 18 13 23.71  & -17 50 40.41  &  19.51  &  16.41  &  14.94  &   9.96  &   9.06  &   8.56  &   8.23  &   8.10  &   7.99  &   8.06  &       keck &     O7-O9   \\
 6  & 18 13 33.82  & -17 51 48.89  &\nodata  &\nodata  &  16.12  &  10.29  &   9.15  &   8.57  &   8.00  &   7.90  &   7.75  &   7.75  &      ukirt &     O6-O7f  \\
 7  & 18 13 22.52  & -17 53 50.14  &  18.63  &  15.60  &  16.57  &  10.30  &   9.27  &   8.66  &   8.03  &   7.70  &   7.49  &   7.30  &      ukirt &     WN7	  \\
 8  & 18 13 14.49  & -17 54 01.77  &  19.91  &  15.47  &  15.71  &   9.86  &   9.12  &   8.75  &   8.43  &   8.33  &   8.31  &   8.31  &      ukirt &     O7-O9   \\
 9  & 18 13 21.91  & -17 51 35.91  &  21.17  &\nodata  &  16.24  &  10.74  &   9.80  &   9.34  &   8.95  &   8.90  &   8.83  &   8.91  &      ukirt &     O7-O9   \\
11  & 18 13 23.62  & -17 53 24.46  &\nodata  &\nodata  &\nodata  &  11.74  &  10.85  &   9.59  &  10.16  &  10.32  &  10.02  &  99.99  &       keck &     B0-B3   \\
12  & 18 13 41.33  & -17 55 41.14  &\nodata  &  17.42  &  18.09  &   9.07  &   7.50  &   6.75  &   6.41  &   6.31  &   5.87  &   5.95  &       keck &     B9-A2   \\
13  & 18 13 29.18  & -17 55 14.55  &  21.09  &\nodata  &  15.66  &   8.79  &   7.56  &   6.85  &   6.80  &   6.15  &   6.03  &   5.99  &       keck &     B0-B3   \\
14  & 18 13 26.47  & -17 52 49.48  &  16.27  &  14.25  &  13.01  &   8.21  &   7.38  &   6.91  &   6.81  &   6.44  &   6.38  &   6.32  &       keck &     B0-B3   \\
15  & 18 13 20.98  & -17 49 46.88  &  19.67  &\nodata  &  15.40  &   8.93  &   7.78  &   6.96  &   6.76  &   5.91  &   5.71  &   5.42  &       keck &     cLBV	  \\
16  & 18 13 19.13  & -17 52 57.43  &\nodata  &\nodata  &  16.90  &   9.43  &   8.10  &   7.33  &   6.94  &   6.47  &   6.40  &   6.19  &       keck &     O8-O9If \\
17  & 18 13 27.83  & -17 52 33.19  &\nodata  &  13.46  &  13.13  &   9.70  &   8.97  &   8.52  &   8.29  &   8.23  &   8.16  &   8.20  &       keck &     B0-B3   \\
18  & 18 13 25.25  & -17 53 55.05  &  20.13  &  16.60  &  14.02  &  10.05  &   9.05  &   8.61  &   8.25  &   8.09  &   7.96  &   8.04  &      ukirt &     O7-O9   \\
19  & 18 13 25.20  & -17 53 55.83  &  19.49  &  16.28  &  16.20  &  10.25  &   9.25  &   8.72  &   8.41  &   8.25  &   8.21  &   8.20  &      ukirt &     O7-O9   \\
20  & 18 13 21.54  & -17 54 35.65  &\nodata  &\nodata  &\nodata  &  10.60  &   9.61  &   9.14  &   8.70  &   8.62  &   8.49  &   8.59  &      ukirt &     O7-O8   \\
21  & 18 13 25.06  & -17 53 09.34  &\nodata  &  16.84  &  13.98  &  10.73  &   9.78  &   9.29  &   9.03  &   8.86  &   8.85  &   8.77  &      ukirt &     O9-B3   \\
22  & 18 13 19.25  & -17 53 22.62  &\nodata  &\nodata  &\nodata  &  11.42  &  10.10  &   9.34  &   8.76  &   8.57  &   8.40  &   8.50  &      ukirt &     O9-B3   \\
23  & 18 13 12.47  & -17 54 28.26  &  19.01  &  15.37  &  13.95  &  10.65  &   9.83  &   9.44  &   9.19  &   9.21  &   9.10  &   9.14  &      ukirt &     O9-B3   \\
24  & 18 13 23.87  & -17 53 10.10  &\nodata  &\nodata  &\nodata  &  11.68  &  10.75  &  10.31  &   9.96  &   9.89  &   9.80  &   9.91  &      ukirt &     O9-B3   \\
25  & 18 13 25.90  & -17 53 57.45  &\nodata  &\nodata  &  20.28  &  12.33  &  11.34  &  10.82  &  10.48  &  10.45  &  10.25  &  10.46  &      ukirt &     O9-B3   \\
26  & 18 13 25.23  & -17 51 35.64  &  15.37  &  14.90  &\nodata  &  13.64  &  11.74  &  10.84  &  10.16  &  10.12  &  10.00  &  10.07  &      ukirt &     O9-B3   \\
\enddata
\tablecomments{For each star, number designations and coordinates
(J2000) are followed by magnitudes measured in different bands. 
\J,\H, and \Ks\ measurements are from 2MASS, while the magnitudes at 3.6 \um,
4.5 \um, 5.8 \um, and 8.0 \um\ are from GLIMPSE.  $B$, $V$, and $R$
associations are taken from the astrometric catalog NOMAD. }
\tablenotetext{a}{2MASS magnitudes are above saturation limits, and have uncertainties $\sim0.3$ mag}
\end{deluxetable}

\subsection{Photometric results}

\subsubsection{Cluster size}
We assume as a cluster center  the flux weighted centroid of a 2MASS \Ks-band
smoothed image (RA=18:13:24.15, DEC=-17:53:29.64). We take as a cluster radius 
the average distance from the cluster center where the surface brightness 
becomes equal to the average brightness of the surrounding field (3.5\arcmin).
The 61 spectroscopically observed stars are located within 6.5\arcmin\ from the 
cluster center, but predominantly (50) within the cluster radius. 
Within 3.5\arcmin\ from the cluster center, we observed  41 stars out of the 48 
stars with  \Ks$<9.5$ mag, and  30 stars out of the 31 stars  with \Ks$<9.0$ mag.  

A fraction of 48\% of stars brighter than \Ks=9.5 mag,  within the cluster radius, have 
early types, and are likely to be members. 
The high degree of completeness of the spectroscopic observations of the 
bright sample allows  for a precise numbering of stars 
in various post-main sequence  phases (WR, cLBV, RSG, and OB stars).

\subsubsection{Color\--Magnitude Diagram and cluster reddening}
\label{sec.CMD}

\begin{figure*}[!]
\begin{center}
\resizebox{0.8\hsize}{!}{\includegraphics[angle=0]{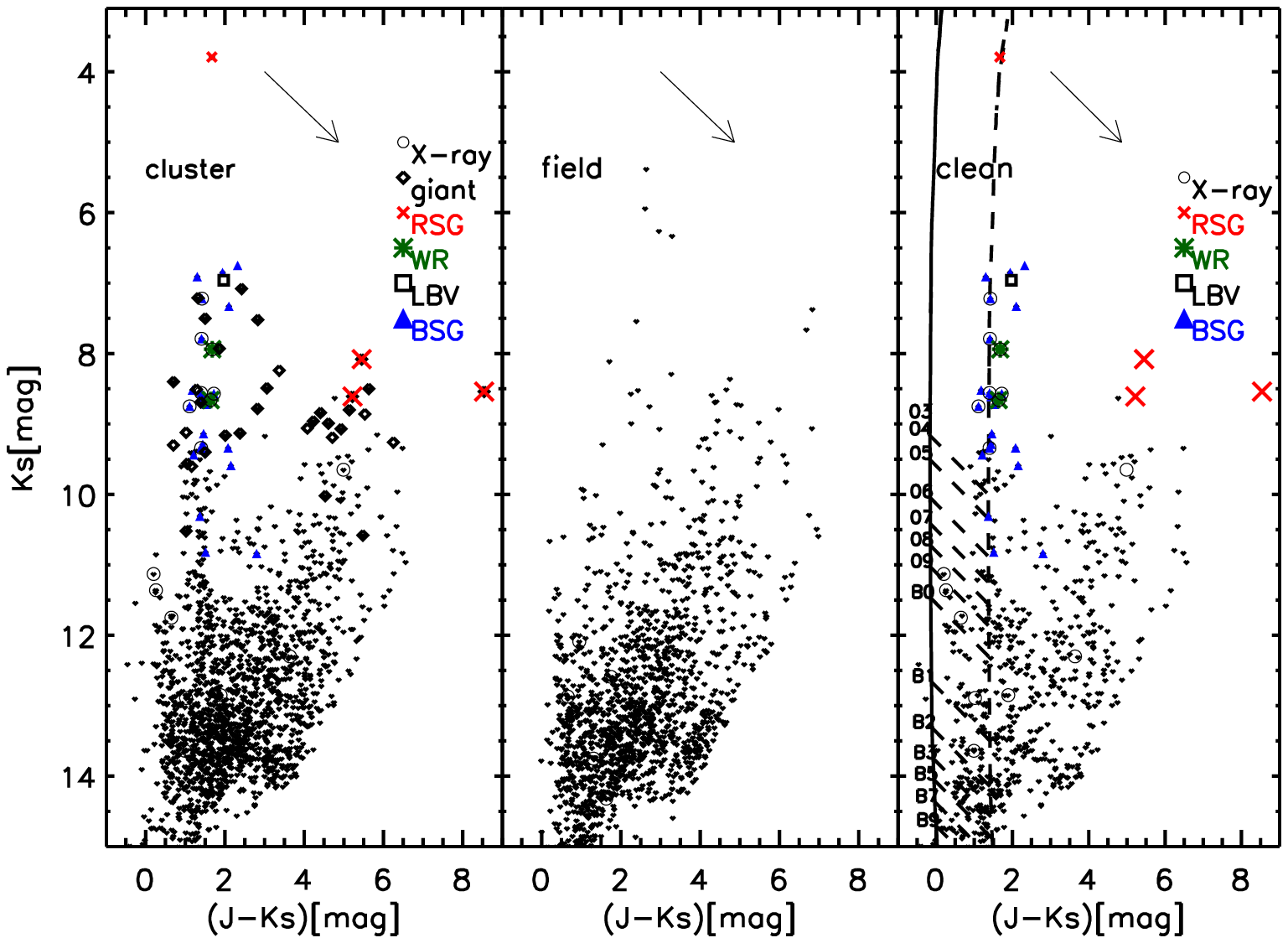}}
\resizebox{0.6\hsize}{!}{\includegraphics[angle=0]{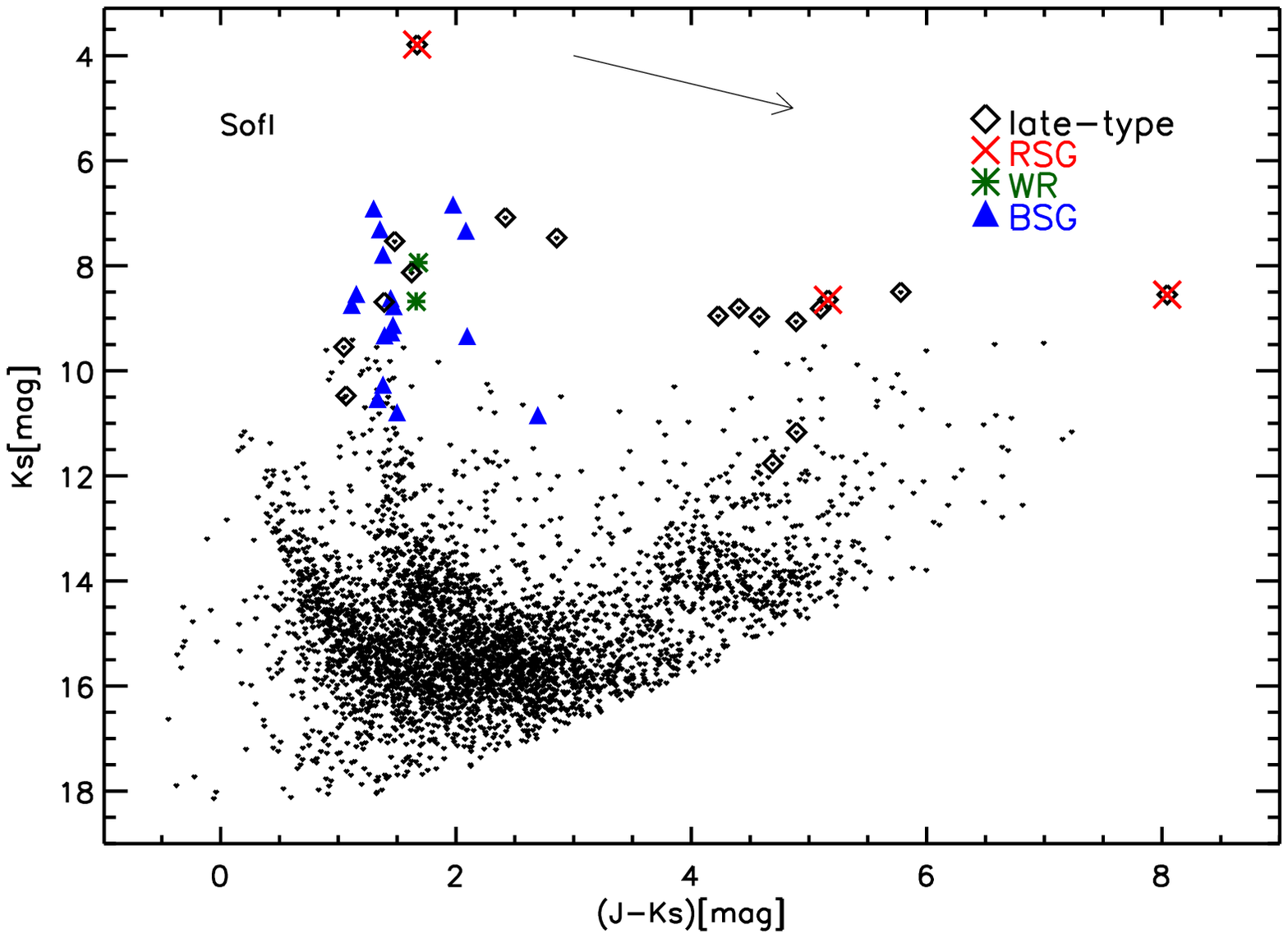}}
\end{center}  
\caption{\label{fig.cmd} {\bf Left upper panel:} A 2MASS $J-$\Ks\ versus \Ks\  diagram of 
point sources detected within 3.5\arcmin\ from the cluster center is displayed. 
{\bf Middle upper panel:} For comparison, A CMD for  stars taken from  a 
surrounding annulus (with a minor radius of 7\arcmin\ and area equal to that of the
cluster) is show. {\bf Right upper panel:} A CMD of the cluster field after statistical
decontamination is shown. The vertical line indicates an isochrone of 4.5 Myr,
with solar metallicity, and with a distance of 4.8 kpc (solid) \citep{lejeune01}; 
the same isochrone  shifted to a reddening of  \Aks=0.83 mag is indicated with a dashed line. 
The locations of OB stars on the ZAMS with \Aks=0 mag 
are marked with labels; dashed lines show their paths to  \Aks=0.83 mag.
The arrow indicates the reddening vector for \Aks=1 mag, based on the extinction law by \citet{messineo05}.
Stars spectroscopically observed are marked with 
special symbols, as listed in the legend. 
{\bf Bottom:} $J-$\Ks\ versus \Ks\ diagram of point sources detected in the SofI images. 
Symbols are the same as in the upper panels.}
\end{figure*}


\begin{figure*}[!]
\begin{center}
\resizebox{0.9\hsize}{!}{\includegraphics[angle=0]{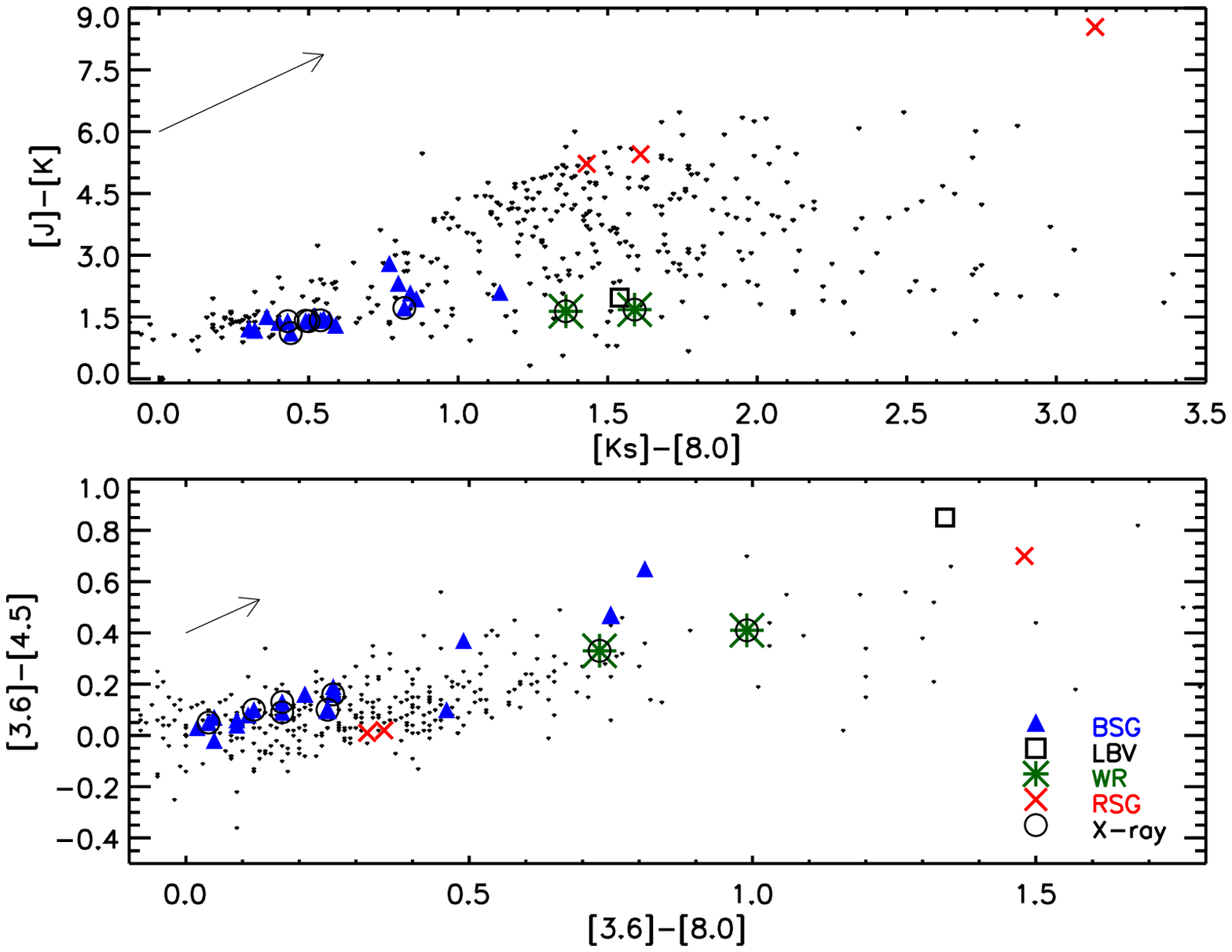}}
\end{center}  
\caption{\label{fig.colcol} {\bf Upper panel:} \Ks$-[8.0]$ versus $J-$\Ks\ diagram of
point sources detected within 3.5\arcmin\ from the cluster center. The arrow
indicates the reddening vector for  \Aks$=1.0$ mag, which is calculated  by following  a 
near-infrared power law with $\alpha=$1.9 \citep{messineo05}, and by using the extinction 
ratios $A_{3.6}$/\Aks, $A_{4.5}$/\Aks, $A_{5.8}$/\Aks, $A_{8.0}$/\Aks of \citet{indebetouw05}.
Symbols are as seen in Fig.\ \ref{fig.cmd}.
{\bf Bottom panel:} GLIMPSE [3.6]-[8.0] versus [3.6]-[5.8] diagram of point sources 
detected within 3.5\arcmin\ from the cluster center. Symbols and arrow are as shown in the top panel.}
\end{figure*}

In the top panels of Fig.\ \ref{fig.cmd}, we display a ($J-$\Ks) vs. \Ks\ diagram of 2MASS 
data points  within the cluster radius of 3.5\arcmin, as well as a diagram of  a comparison field.  
The same CMD with deeper SofI data, covering the central  4.9\arcmin$\times$4.9\arcmin,
is shown in the bottom panel of Fig.\  \ref{fig.cmd}.

Several distinct populations of stars are seen in the CMD. There is a  sequence of stars
at $J-$\Ks=0.5 mag, which we attribute to field stars in the closer 
Sagittarius-Carina spiral arm.
There is a broad sequence  with $J-$\Ks\ from 3 to 6 mag,  which is populated mainly
by  field giant stars in the background of the cluster. Three candidate RSGs were detected
from this sequence. Finally, there  is a sequence of bright stars  with $J-$\Ks$=\sim 1.5$ mag
and with \Ks\ from 3 to 14 
mag     (bottom panel of Fig.\  \ref{fig.cmd}). This sequence is seen 
only in the  cluster field, and this  is the brightest extension of the cluster sequence. 

To confirm and isolate the cluster sequence, we performed a statistical decontamination.  
2MASS data were used in order to have a comparison field. 
We counted the number of stars in a grid   of 0.5 mag in ($J-$\Ks) and of 1.0 mag in \Ks.
From each bin of the cluster CMD, we subtracted a number of stars equal to that of the 
corresponding field bin. First, we subtracted all contaminating giant stars (spectroscopically
detected), then we continued with a random subtraction. The decontaminated diagram is 
displayed in the right panel of Fig.\ \ref{fig.cmd}, and different symbols indicate
known spectral types. The cluster sequence  appears  populated by massive stars.  
An average interstellar extinction of  \Aks$=0.83\pm0.2$ mag (Av$=9.1$ mag) was estimated 
by matching the colors of the observed cluster sequence with a theoretical isochrone of solar
composition from the Geneva group \citep{lejeune01}, and by assuming a power--law
extinction curve {\it \Al} $\propto \lambda^{-1.9}$ \citep{messineo05}. This
measurement is independent of age since the isochrones are  almost vertical sequences in this
plane. An isochrone of 4.5 Myr and solar metallicity, taken from the
non-rotating models of the Geneva group, is over-plotted on the diagram \citep{lejeune01}.

\section{Luminosities of the massive stars}

In the following, we  analyze the properties (colors, magnitudes, and luminosities) of massive stars  
in the Cl 1813-178 cluster, which we list in Table \ref{table.properties}.  A kinematic distance of 
4.8 kpc is used (see below the RSG section).
For each star, we assumed an intrinsic color consistent with its stellar  spectral type
\citep{koornneef83,martins06,wegner94}; we used the extinction law by \citet{messineo05}, and the 
kinematic distance, and we estimated a value of \Aks\ and  \Mk\  (see Table \ref{table.properties}). The 
\Mk\ values were  transformed into bolometric magnitudes, Mbol, by adding the bolometric corrections.
The adopted effective temperatures and  bolometric corrections per spectral type 
are given in the Appendix.

\subsection{Red supergiants}
\label{sec.rsg}

Star \#1 is the brightest star within the cluster area, \Ks=3.8 mag, and is consistent with being 
a K2-K5 I cluster member (PaperI).
From the radial velocity ($V_{LSR}=62 \pm 4$) of star \#1 (PaperI), and the Galactic rotation 
curve by \citet{reid09}, we obtained a revised near kinematic distance of $4.8^{+0.25}_{-0.28}$ kpc.
With intrinsic colors from \citet{koornneef83}, bolometric correction, \BCK, taken from 
\citet{levesque05}, we estimated a luminosity  \Lstar$= 9\pm7 \times 10^4$ \Lsun.

Stars \#32, \#38, and  \#39 have EW(CO)s larger than the typical values for red giants. 
Stars \#32 and \#38 were observed with NIRSPEC, and their $K$-band spectral coverage 
allows us to study both the CO band heads, and the shape of the continuum. The large 
values of EW(CO)s and the absence of water absorption suggest a supergiant luminosity class
for stars \#32 and \#38 \citep[][]{comeron04}.
Star \#39  was observed with UIST on UKIRT, with the $K-$long filter only, and we do not have 
information on the shape of its continuum. We derived  spectral types M2.5I, M3.5I, and M1I, 
respectively, by assuming a supergiant luminosity class for all three stars. 

Stars \#32, \#38, and  \#39 are among the brightest stars with red $J-$\Ks\ (from 3 to 9 mag), 
much redder than the cluster sequence as shown in Figs.\  \ref{fig.cmd},  and \ref{fig.colcol}.
By assuming  intrinsic $H-$\Ks\ and $J-$\Ks\ colors for a given spectral type (see the Appendix), 
we obtained \Aks = 2.3 mag, 3.6 mag, and 2.3 mag,  respectively. 
These values of interstellar extinction are the combined contribution of interstellar and possible 
circumstellar extinction \citep{messineo05}. 
By analyzing the distribution of detected stars in the $Ks-[8.0]$ vs $J-$\Ks\ plane (\ref{fig.colcol}),
star \#38 appears to show an excess in $Ks-[8.0]$,  which is suggestive of a circumstellar envelope. 
Stars \#32, and  \#39 have  $Ks-[8.0]$  and $J-$\Ks\ colors 
in agreement within errors with those of naked late-type stars \citep{messineo04,messineo05}.
We corrected for the total extinction by using the intrinsic colors for  given spectral types
and the extinction law by \citet{messineo05}.
We then artificially reddened the stars to emulate the average cluster condition.
By assuming these stars were moved along  the reddening vector to the average cluster 
extinction (\Aks=0.8 mag),  we obtained  \Ks=6.7 mag, 5.8 mag, and 7.2 mag, respectively. 

If these RSGs were at the same distance as the stellar cluster they should be 
the brightest stars  at near-infrared wavelengths, and  their \Ks\ magnitudes
would increase with  later spectral types. Their  spectral types are later than  the bright
(\Ks=3.8 mag) K2I star,  and their  de-reddened magnitudes are similar to those of the brightest 
OB stars (\Ks $=\sim7$ mag). The spectral types and magnitudes of these three RSGs are not  
consistent with cluster membership; they  are too faint to be RSGs and part of the cluster.
This reasoning is valid also for stars with moderate mass-loss, i.e. \#38, because
bolometric corrections \BCK\  are almost constant for colors \Ks$-[8.0] <3$ mag  
\citep[chapter5, Figs. A1 and A2,][]{messineothesis}.

We estimated a density of non-member RSGs equal to 0.16 in the central 3.5\arcmin.
These RSGs could represent an outburst farther away, perhaps related to the Galactic bar.

\subsection{Wolf Rayet stars}
Star \#4, a WN7b, was  spectroscopically detected by \citet{hadfield07}.
Hadfield et al. performed a successful  color-based selection of candidate WR stars 
with 2MASS and GLIMPSE data.
WRs have an infrared excess (see Figs.\ \ref{fig.colcol} and \ref{fig.cmd}), which is 
caused by free-free emission, and do not follow  the reddening vector.
Star \#7 is a newly detected WN7o star, its colors fall  within the Hadfield et al.\ criteria.

For stars \#4 and \#7, we used intrinsic colors  for WN stars with broad and narrow lines, respectively, 
derived by \citet{crowther06} (see the Appendix).
We obtained an interstellar extinction \Aks=0.7 mag, and \Aks=0.8 mag, respectively.
By adopting  the kinematic distance,    \BCK\ from \citet{crowther06}, we obtained \Lstar 
$= 5.8\pm2.5 \times 10^5$ \Lsun\ for star \#4, and \Lstar $=4.8\pm 2.1 \times 10^5$ \Lsun\ for star \#7.

\subsection{Luminous blue variables} 

Among the observed spectra, we detected one candidate cLBV star (\#15).
LBV stars are rare massive supergiants  in transition  towards the Wolf-Rayet phase
\citep[e.g.][]{conti95,nota95,figer97,conti84}.
Since their evolutionary paths on the CMD are quite uncertain, 
their detections in stellar clusters  are of primary  importance. 
The K-band spectrum of star \#15 is similar to that of a P Cygni-type B supergiant, 
cLBV, e.g. the [OMN2000]LS1 star presented by \citet{clark09}. 
We used CMFGEN, the iterative non-LTE line blanketing method presented by 
\citet{hillier98}, to estimate the physical properties of this supergiant.
For details on the modeling see \citet{najarro99}, \citet{najarro01} 
and \citet{clark09}.

The K-band spectra provided the primary diagnostics (the HeI lines at 2.058 \um\ and 2.112 \um,
and the MgII lines at 2.138 \um\ and 2.144 \um). We obtained  an effective temperature 
T$_{eff}= \sim 16,000\pm300$ k, and a value of \Lstar$= 2.1\pm1.5 \times 10^5$ \Lsun. This value matches
the luminosities of  the faintest P-Cygni supergiants known in the Milky Way \citep{clark05,clark09}.

\subsection{OB stars} 
\label{sec.OB}

We detected  21 OB stars  in the Cl 1813-178 cluster. The current dataset of medium-resolution
K-band spectra  allow for a spectral classification   of OB stars typically within $\pm$2
spectral types \citep{hanson96}. The combination of spectral and  photometric properties suggests
that we have detected 14 OB supergiants, 4 OB giants, and  3 OB main sequence stars. 

For star \#6 (O6-O7If+), when using    an intrinsic ($H-$\Ks)$_0=-0.1$ mag,
and  \BCK=$-4.01$ mag \citep{martins06}, we obtained  \Mk=$-$5.87 mag and  \Lstar$=8.2\pm2.3
\times 10^5$ \Lsun.  The estimated value of \Mk\ is consistent with the value given for an O6-O7
star by \citet{martins06} and \citet{clark05}.

Star \#16 is another rare O supergiant, an O8-O9If type. By assuming  ($H-$\Ks)$_0=-0.1$ mag, 
and \BCK=$-3.84$ mag \citep{martins06}, we obtained \Mk=$-7.32$ mag and 
\Lstar$= 23\pm6 \times 10^5$  \Lsun. 
Star \#16 is the brightest early-type cluster member, and it is located close to the observed Humphreys-Davidson 
limit for stars with similar effective temperatures \citep[][]{clark05}.

From a comparison with the \Mk\ values by \citet{martins06} and \citet{panagia73}, 
the OB stars \#25, \#24, and \#26 are likely dwarfs, since they have \Mk$\le -4 $ mag 
(\Lstar$\le 1.4\times10^5$ \Lsun).
The OB stars  \#9,  \#21, \#22, and \#23 are probably of luminosity classes III or II,
since they  are fainter than a typical BSG. Their \Mk\ values range from $-$4.7 to $-$5.36 mag
(\Lstar\ varies from $1.1\times10^5$ to  $1.5\times10^5$ \Lsun).
All remaining OB stars are supergiants.

The large spread in  magnitudes  of OB supergiants ($\sim 3$ mag in \Ks-band) is not
surprising.  A similar observational spread is observed in the 2MASS \Ks\ magnitudes of
BSGs in Westerlund 1 \citep[][]{clark05}.

{\tiny
\begin{deluxetable}{rrrrrrrrrrll}
\rotate
\tablewidth{0pt}
\tablecaption{\label{table.properties} Derived properties of stellar members.}
\tablehead{
\colhead{ID}& 
\colhead{Ks}& 
\colhead{J-Ks}& 
\colhead{H-Ks}& 
\colhead{E$_{J-Ks}$}& 
\colhead{E$_{H-Ks}$}& 
\colhead{A$_{K}$} &
\colhead{M$_{K}$}&
\colhead{Lum}&
\colhead{log(Teff[K])}&
\colhead{Sp. type} &
\colhead{Class}
}
\startdata
 1  &   3.79  &   1.67 &   0.44  &   1.02  &   0.31 &   0.54$\pm$  0.13 & -10.16  &   4.95 $^{ 0.24}_{-0.25}$ &   3.60 $^{ 0.01}_{-0.02}$     & RSG  &I \\
 2  &   7.22  &   1.42 &   0.50  &   1.50  &   0.54 &   0.80$\pm$  0.02 &  -6.99  &   5.70 $^{ 0.34}_{-0.34}$ &   4.31 $^{ 0.12}_{-0.12}$     & B0B3  &I \\
 3  &   7.79  &   1.41 &   0.46  &   1.49  &   0.50 &   0.79$\pm$  0.02 &  -6.41  &   5.46 $^{ 0.34}_{-0.34}$ &   4.31 $^{ 0.12}_{-0.12}$     & B0B3  &I \\
 4  &   7.94  &   1.68 &   0.66  &   1.31  &   0.39 &   0.69$\pm$  0.05 &  -6.16  &   5.76 $^{ 0.16}_{-0.17}$ &   4.70 $^{ 0.04}_{-0.05}$     & WN7  &I \\
 5  &   8.56  &   1.40 &   0.50  &   1.61  &   0.60 &   0.87$\pm$  0.02 &  -5.72  &   5.75 $^{ 0.12}_{-0.13}$ &   4.50 $^{ 0.03}_{-0.03}$     & O7O9  &I \\
 6  &   8.57  &   1.72 &   0.58  &   1.93  &   0.68 &   1.03$\pm$  0.02 &  -5.87  &   5.92 $^{ 0.11}_{-0.12}$ &   4.54 $^{ 0.02}_{-0.02}$     & O6O7If  &I \\
 7  &   8.66  &   1.64 &   0.61  &   1.51  &   0.50 &   0.80$\pm$  0.05 &  -5.55  &   5.68 $^{ 0.16}_{-0.17}$ &   4.70 $^{ 0.04}_{-0.05}$     & WN7  &I \\
 8  &   8.75  &   1.11 &   0.37  &   1.32  &   0.47 &   0.71$\pm$  0.01 &  -5.36  &   5.61 $^{ 0.12}_{-0.13}$ &   4.50 $^{ 0.03}_{-0.03}$     & O7O9  &I \\
 9  &   9.34  &   1.40 &   0.46  &   1.61  &   0.56 &   0.86$\pm$  0.02 &  -4.93  &   5.47 $^{ 0.13}_{-0.14}$ &   4.51 $^{ 0.03}_{-0.03}$     & O7O9  &III \\
11  &   9.59  &   2.15 &   1.26  &   2.23  &   1.30 &   1.37$\pm$  0.02 &  -5.19  &   4.98 $^{ 0.34}_{-0.34}$ &   4.31 $^{ 0.12}_{-0.12}$     & B0B3  &III \\
12  &   6.75  &   2.32 &   0.75  &   2.23  &   0.74 &   1.19$\pm$  0.02 &  -7.85  &   5.07 $^{ 0.13}_{-0.14}$ &   3.98 $^{ 0.03}_{-0.03}$     & B9A2  &I \\
13  &   6.85  &   1.94 &   0.71  &   2.02  &   0.75 &   1.08$\pm$  0.03 &  -7.64  &   5.96 $^{ 0.34}_{-0.34}$ &   4.31 $^{ 0.12}_{-0.12}$     & B0B3  &I \\
14  &   6.91  &   1.30 &   0.47  &   1.38  &   0.51 &   0.74$\pm$  0.02 &  -7.24  &   5.79 $^{ 0.34}_{-0.34}$ &   4.31 $^{ 0.12}_{-0.12}$     & B0B3  &I \\
15  &   6.96  &   1.97 &   0.82  &   1.99  &   0.84 &   1.09$\pm$  0.02 &  -7.53  &   5.38 $^{ 0.22}_{-0.23}$ &   4.20 $^{ 0.01}_{-0.01}$     & LBV  &I \\
16  &   7.33  &   2.10 &   0.77  &   2.31  &   0.87 &   1.24$\pm$  0.01 &  -7.32  &   6.36 $^{ 0.11}_{-0.12}$ &   4.48 $^{ 0.01}_{-0.01}$     & O8O9If  &I \\
17  &   8.52  &   1.18 &   0.45  &   1.26  &   0.49 &   0.68$\pm$  0.03 &  -5.56  &   5.13 $^{ 0.34}_{-0.34}$ &   4.31 $^{ 0.12}_{-0.12}$     & B0B3  &I \\
18  &   8.61  &   1.44 &   0.44  &   1.65  &   0.54 &   0.88$\pm$  0.05 &  -5.67  &   5.74 $^{ 0.13}_{-0.14}$ &   4.50 $^{ 0.03}_{-0.03}$     & O7O9  &I \\
19  &   8.72  &   1.53 &   0.53  &   1.74  &   0.63 &   0.94$\pm$  0.02 &  -5.62  &   5.72 $^{ 0.12}_{-0.13}$ &   4.50 $^{ 0.03}_{-0.03}$     & O7O9  &I \\
20  &   9.14  &   1.46 &   0.47  &   1.62  &   0.51 &   0.86$\pm$  0.03 &  -5.13  &   5.58 $^{ 0.11}_{-0.12}$ &   4.54 $^{ 0.01}_{-0.01}$     & O7O8  &I \\
21  &   9.29  &   1.44 &   0.49  &   1.60  &   0.57 &   0.86$\pm$  0.03 &  -4.98  &   5.04 $^{ 0.38}_{-0.38}$ &   4.36 $^{ 0.13}_{-0.13}$     & O9B3  &III \\
22  &   9.34  &   2.08 &   0.76  &   2.24  &   0.84 &   1.22$\pm$  0.03 &  -5.28  &   5.16 $^{ 0.38}_{-0.39}$ &   4.36 $^{ 0.13}_{-0.13}$     & O9B3  &III \\
23  &   9.44  &   1.21 &   0.39  &   1.34  &   0.42 &   0.71$\pm$  0.02 &  -4.67  &   5.03 $^{ 0.30}_{-0.30}$ &   4.39 $^{ 0.11}_{-0.11}$     & O9B3  &III \\
24  &  10.31  &   1.37 &   0.44  &   1.50  &   0.47 &   0.79$\pm$  0.02 &  -3.89  &   4.72 $^{ 0.30}_{-0.30}$ &   4.39 $^{ 0.11}_{-0.11}$     & O9B3  &V \\
25  &  10.82  &   1.51 &   0.52  &   1.64  &   0.55 &   0.88$\pm$  0.04 &  -3.46  &   4.55 $^{ 0.30}_{-0.30}$ &   4.39 $^{ 0.11}_{-0.11}$     & O9B3  &V \\
26  &  10.84  &   2.80 &   0.90  &   2.93  &   0.93 &   1.55$\pm$  0.02 &  -4.12  &   4.81 $^{ 0.30}_{-0.30}$ &   4.39 $^{ 0.11}_{-0.11}$     & O9B3  &V \\
\enddata
\end{deluxetable}
}

\subsection{X--ray emitters}

\begin{deluxetable}{rlrrrrrrr}
\tablecaption{\label{table.xray} Massive stars with X-ray emission.}
\tablehead{
\colhead{ID}& 
\colhead{\Ks }& 
\colhead{Sp. Type}& 
\colhead{Id$_{Ch}$} &
\colhead{Counts$_{Ch}$}&
\colhead{HR}&
\colhead{Id$_{XMM}$}&
\colhead{Counts$_{XMM}$} &
\colhead{Lx }
}
\startdata
    &  [mag] &      &    &         &         &         &          &  [$10^{31}$ erg s$^{-1}$] \\   
\hline
 2  &  7.22  &  OB  &39  &  14.10  &$-$0.33  & \nodata &   \nodata& 4.1\\
 3  &  7.79  &  OB  &43  &  16.40  &$-$0.53  & \nodata &   \nodata& 4.8\\
 4  &  7.94  &  WR  &24  & 272.40  &   0.78  &  2      &       238&80.5\\
 5  &  8.56  &  OB  &41  & 214.50  &   0.42  &  4      &       138&63.4\\
 6  &  8.57  &  OB  &58  &  13.60  &$-$0.57  & \nodata &   \nodata& 4.0\\
 7  &  8.66  &  WR  &37  &  34.90  &   0.89  & \nodata &   \nodata&10.3\\
 8  &  8.75  &  OB  &27  &  13.00  &$-$0.82  & \nodata &   \nodata& 3.8\\
 9  &  9.34  &  OB  &36  &   9.50  &$-$0.20  & \nodata &   \nodata& 2.8\\
10  &  9.60  &  K   &71  &  71.10  &$-$0.31  & \nodata &   \nodata&21.0\\
\enddata
\tablecomments{For each star, number designations,  \Ks\ magnitudes, and spectral types
from Tables \ref{table.members} and \ref{table.nonmembers} are followed by the number 
designations (ID$_{Ch}$), counts (counts), and hardness
(HR) from the Chandra observations by \citet{helfand07}, and by the number designations (Id$_{XMM}$)
and XMM counts reported by \citet{funk07}. Estimates of \Lx\ are taken from PaperI,
and assume a distance of 4.7 kpc,
N(H)=$1.6 \times 10^{22}$ cm$^{-2}$, a power-law model, and a photon index of 1.5.
X--ray luminosities are given in units of $10^{31}$ erg s$^{-1}$.}
\end{deluxetable}

X-ray emission  may be generated in the circumstellar envelopes of massive stars
due to their strong shocked winds  \citep{lucy80}. 
Single OB stars    emit with a typical X-ray luminosity of \Lx= 10$^{31-33}$ erg
s$^{-1}$ \citep{pollock87}, and have a typical ratio 
between the X-ray and bolometric luminosities of about $ 10^{-7}$. 
OB+OB and OB+WR binaries generally have higher luminosities
\citep[\Lx$= 10^{32-35}$ erg s$^{-1}$,][]{clark08}.   X-ray emission enables us to
characterize the physical condition of stellar atmospheres, and to identify 
binary systems.
 
X-ray observations of the Cl 1813-178 cluster region  were performed by \citet{funk07}
and \citet{helfand07}, successfully detecting  a large number (75) of X-ray emitters.
We looked for possible associations between X-ray emitters and  
massive members of the Cl 1813-178 cluster (PaperI).
All but one X-ray sources with a bright 2MASS counterpart (\Ks$<9.6$ mag), were found
associated with early-type cluster members (see Table \ref{table.xray}).
Two X-ray emitters coincide  with WR stars (\#4 and \#7);  six others are associated  
with OB stars (\#2, \#3, \#5, \#6, \#8, and  \#9).  The remaining
X-ray emitter  \citep[\#71,][]{helfand07}  coincides  with  star \#10,  
which is likely a cluster non-member. This star  is located  6.6\arcmin\  from the cluster center, 
outside of the cluster radius (3.5\arcmin).   Its spectrum has CO bands at 2.29 \um, which indicate a 
late-type star. 

We estimated the ratios between the X-ray and bolometric luminosities,
and we compared the ratio, hardness  and X-ray luminosities of our nine X-ray emitters with 
those of Chandra point sources detected in  Westerlund 1 \citep[Fig.\ 5][]{clark05}.
In Westerlund 1, WR stars have luminosities \Lx\ larger than $10^{32-34}$ erg s$^{-1}$ and 
hardness from $-0.1$ to 1, while most of the detected OB stars have  \Lx=10$^{32}$  erg s$^{-1}$ 
(which is consistent with a ratio of $\sim 10^{-7}$) and hardness between $-0.8$ and $-0.1$,
as expected for single stars with shocked winds.
Clark et al. suggest that all  OB stars with  X-ray emission significantly harder than $-$0.5 
in Westerlund 1 are binary systems.  This conclusion is supported by 
emerging evidence of a high-binary fraction of massive stars in Westerlund 1 \citep{bonanos07}.
\citet{skinner10} propose a number of other possible scenarios to explain the
X-ray emission of single WN stars. Magnetic wind confinement could also explain the presence of 
a hot plasma component without invoking the presence of a close companion. However, current
detections of magnetic fields in WN stars are absent.

Star \#4, a WN7 star, is the  brightest X-ray emitter. It is associated with the Chandra source
\#24  \citep{helfand07}, and coincides also with the XMM source \#2 of \citet{funk07}. The ratio 
between the X-ray and bolometric luminosities of star \#4  is $3.7 \times 10^{-7}$.  For  star \#7, 
we obtained a ratio 7 times fainter. Both  ratios are consistent with  those measured in  other
WN stars by \citet{skinner10}.    Their high values of hardness (0.78 and 0.89 in Table
\ref{table.xray}) are consistent with those of   colliding wind binaries \citep{clark08}. 
Since only a few other late WN stars have been detected  in X--ray \citep{skinner10}, our new detections 
are a significant addition.

Star \#5 (O7-O9) was detected by both the XMM and Chandra satellites (see Table
\ref{table.xray}). It is a strong X-ray emitter  (\Lx$=1 \times 10^{32}$ erg s$^{-1}$); the
ratio between the X--ray and bolometric  luminosities is $2.9 \times 10^{-7}$.
Besides the two WR stars, star \#5 is the only other source with a strong
X-ray hardness (0.4). The high values of X-ray luminosity and hardness indicate that  \#5 is
another binary.

For the BSGs \#2 and \#3 (B0-B3), we measured  ratios between the X-ray and bolometric 
luminosities of $2.1-4.3 \times 10^{-8}$, which are in agreement with the ratios measured for single stars 
later than B1 \citep{cohen96,waldron07}. 
Stars \#8 and \#9 (O7-O9) have also a  ratio of about  $2.5 \times 10^{-8}$. 

For star \#6, an O6O7If,  we estimated a ratio of $1.3 \times 10^{-8}$.  The low ratio
 and low hardness ($-$0.57) suggest radiative  shocks in stellar winds.


\section{Spectro-photometric distances}

Massive evolved stars in a young stellar cluster  span a range of masses (see Fig.\
\ref{figure.lejeune}),  therefore, of luminosities. Spectro-photometric distances from OB
supergiants are  less accurate than those from dwarfs and giants. 
When considering  the  dwarfs and the  \Mk\ magnitudes  for  O9, B0, B1, and B2 
dwarfs  \citep{martins06, humphreys84, koornneef83}, we estimated  distances of 
$3.8\pm0.6$ kpc, $3.2\pm0.5$ kpc,   $2.6\pm0.4$ kpc, and $1.9\pm0.3$, respectively. 

From the photometric  properties of the candidate giants (\#9, \#21, \#22, and \#23), and  
\Mk\ values for classes II and III from \citet{humphreys84} and \citet{wegner94}, 
we derived an average spectro-photometric distance of $3.0\pm0.7$ kpc for  class III, 
or $4.5\pm0.6$ for class II. 

For the two WN7 stars, we adopted intrinsic magnitudes from \citet{crowther06}, and the  extinction
law by \citet{messineo05}, and obtained   a distance  of $2.6$ kpc for star  \#4, and   $5.8$ kpc
for star \#7.   The average distance for the two WN7 stars is  $4.2\pm1.6$ kpc.  

The derived spectrophotometric distances are listed in Table \ref{table.distance}.
While the distance estimates from giants and WRs
within errors are consistent with the kinematic distance,
the distance estimates from dwarfs  are only consistent with the kinematic distance if the 
dwarfs are all late O stars.
Higher-resolution spectra are needed to refine the spectral types.

\begin{deluxetable}{lll}
\tablewidth{0pt}
\tablecaption{\label{table.distance} Average spectrophotometric distances}
\tablehead{
\colhead{Sp. type}& 
\colhead{distance}& 
}
\startdata
OB V       &  $2.9\pm0.8$\\
OB III/II  &  $3.8\pm1.0$\\
WN7        &  $4.2\pm1.6$\\
\hline
average    & $3.6\pm0.7$\\
\enddata
\end{deluxetable}

Observations of radio hydrogen recombination lines of the W33 complex reveal two velocity components 
\citep{bieging78,goss78}.  
By using  the radial velocity component at  35 \kms,  and the rotation curve of
\citet{reid09}, we calculated a near  kinematic distance of   $3.53^{+0.40}_{-0.46}$ kpc, while   we
obtained a distance   of  $4.82^{+0.25}_{-0.28}$ kpc with the radial velocity component at 62 \kms. 
A radio monitoring program of methanol masers in the direction of the W33 complex is currently ongoing.  
It will yield parallactic distances of the masers (Brunthaler et al. in preparation).

\section{Progenitor masses and the cluster age}
\label{sec.age}

\begin{figure}[!]
\begin{centering}
\resizebox{1.0\hsize}{!}{\includegraphics[angle=0]{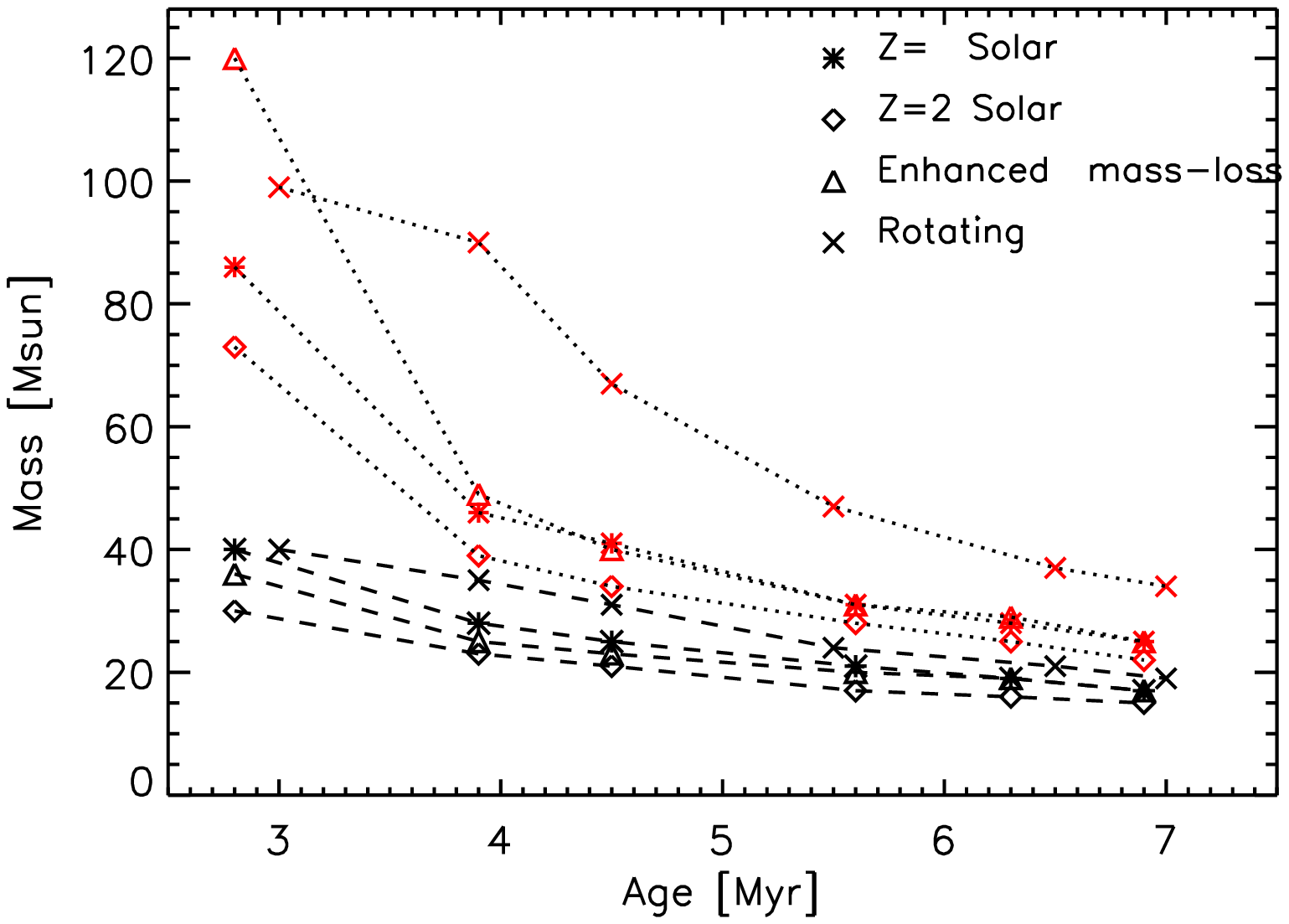}}
\end{centering}
\caption{\label{figure.lejeune} 
For a given age, the mass at the TO  and the maximum mass 
predicted by models are shown.  
Several models are used: a  model by \citet{schaller92} 
with  a solar metallicity
(asterisks), models with metallicity twice solar by \citet{scharer93} (diamonds), 
models with high 
mass-loss by \citet{meynet94} (triangles), and the rotating models by 
\citet{meynet00} (crosses). For each model, the maximum masses are connected with a dotted
line, while the masses at the TO are connected with a dashed line.}
\end{figure}

\begin{figure}[!]
\begin{centering}
\resizebox{1.0\hsize}{!}{\includegraphics[angle=0]{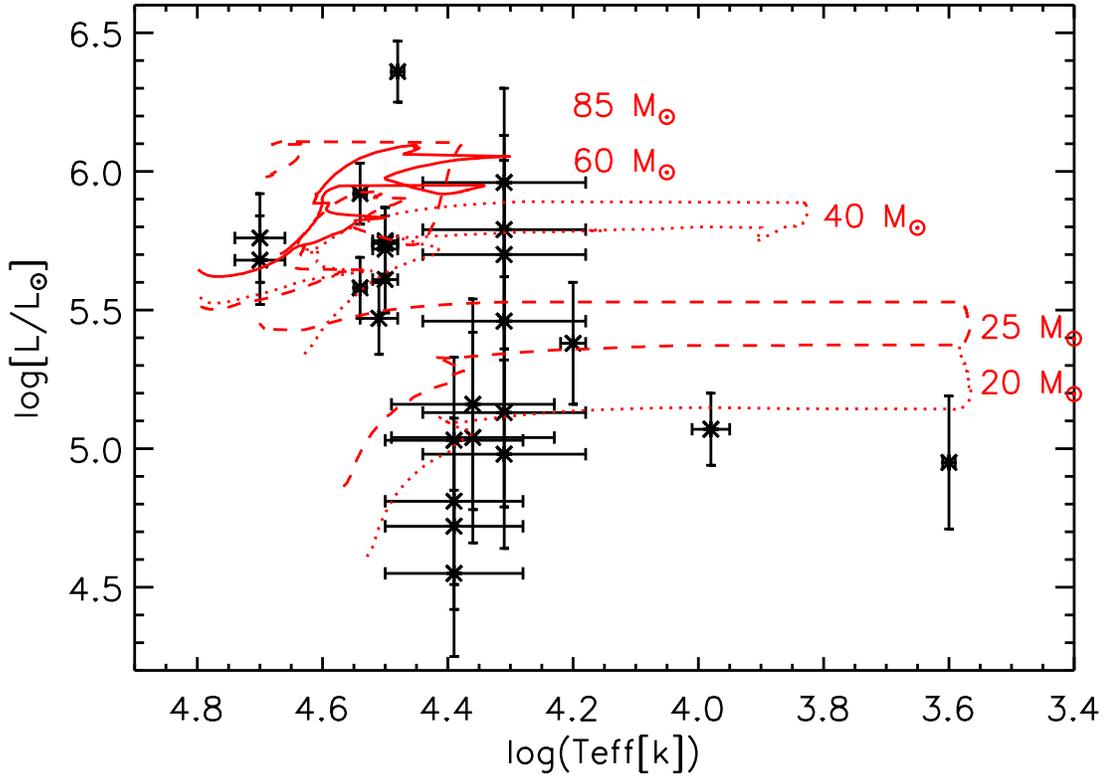}}
\end{centering}
\caption{\label{fig.lum} Luminosities of massive stars 
in the Cl 1813-178 cluster are plotted versus effective temperatures.
Values are taken from Table \ref{table.properties}.
Stellar tracks for stars of 20, 25, 40, 60, and 85 \Msun, 
based on rotating Geneva models with  a solar metallicity, canonical mass-loss rates, and initial 
rotational velocity of 300 \kms\ are also shown with dotted, dashed and continuous lines.
For clarity, we  use different line styles together with labels to indicate each model.
}
\end{figure}

The Cl 1813-178 cluster contains a large number of evolved massive stars (BSGs, a cLBV,
two WN7 stars, and one RSG star), which are listed in Table \ref{table.members}. 
We plot the inferred stellar luminosities versus stellar effective temperatures
in Fig.\  \ref{fig.lum}, together with  theoretical stellar models 
from the Geneva group \citep{meynet00}. By comparison with the models, 
we estimated initial masses from 20 to 100 \Msun.

Models predict a large span of stellar  masses in post main sequence phase, e.g. 
a population of 4.0 Myr would have  a mass of 25-35 \Msun\ at a TO, but it would still contain stars 
of about 50\Msun, or even 100 if rotating models are considered. This is well illustrated in Figure  
\ref{figure.lejeune}, where we show
the predicted maximum initial masses and TO masses as a function of cluster age. 
We used  non-rotating models by \citet{schaller92} and
\citet{scharer93}, models with high-mass loss by \citet{meynet94}, and the rotating models 
by \citet{meynet00}.  Metallicity  ranges from solar to twice solar.
For the same age, different models predict variations in the initial masses of up to 50\%.

The mixture of evolved stars indicates progenitor  masses larger than 20 \Msun,  and more likely  
of 25-60 \Msun, with a few exceptions.

The Of stars are among the most luminous stars. 
The luminosity of the  O8-O9If star suggests an extremely massive star  \citep[100
\Msun,][]{meynet03}. The O6-O7If star has an estimated \Lstar $=8.1\pm2.2 \times 10^5$ \Lsun, 
which is predicted for a 30-70 \Msun.   

WR stars of WN7 type  have been found only  in  stellar clusters with masses at the TO larger than
35-40 \Msun \citep{massey01,clark05}.  
The lumimosities of the WR stars in the Cl 1813-178 cluster  (\Lstar $= 5.2\pm 2.3 \times10^5$ and  
$4.5\pm1.9 \times10^5$ \Lsun) are similar to those inferred for WN7 stars in the Westerlund 1 cluster 
(\Lstar\ from $ 3.1\times10^5$ to  $5.0\times10^5$ \Lsun). The  stellar luminosities suggest  initial 
masses from 40 to 70 \Msun. 

The cLBV and RSG have luminosities expected for  less massive stars ($\sim 20$ \Msun).
However, the 2MASS magnitudes of the RSG star have errors of about 0.3 mag, due to saturation,
and the resulting luminosity could be underestimated.

Models of  young simple stellar populations by \citet{meynet03} predict that stars with masses between
9 to 25-35\Msun\ have a RSG phase. Single WR stars have initial masses greater than $26-30$ \Msun,
while binary WR stars have  initial masses greater than  $20-25$ \Msun\ \citep{eldridge08}.  By assuming
coevality  between the RSG and the WR stars,  the Cl 1813-178 cluster would be between  4 and 6 Myr
years old.   A cluster  with stars exceeding 100 \Msun\ (like the O8-O9If)  would require an
age of 3-4 Myr, and a turn-off (TO) at 35 \Msun.
Therefore, the simultaneous presence of RSGs, WRs, and Of stars, would further narrow the possible
age range to 4-4.5 Myrs. 
However, either the luminosities of the cLBV,  of the RSG star, and/or of the late
B supergiant are  underestimated by $\sim 0.7$ dex, or  the 'cluster' is more of an 'association' with
some degree of  non-coevality. 

The  cLBV star appears to be rather faint. The estimated luminosity of 
$2.4 \times 10^5$ \Lsun\ is similar to that of the HD168607 and HD316285
LBVs \citep{clark05}. Other known LBVs in clusters are
typically among the most luminous members, e.g. qF362 and the pistol 
star in Quintuplet \citep[e.g.][]{mauerhan10}.
In Westerlund 1 the W243 LBV  has an initial mass of about 40 \Msun,
which is consistently similar to  the masses of  the cluster WRs (40-50 \Msun), and
larger than the estimated mass at the TO (30 \Msun) \citep{ritchie09}.
The cLBV in Cl1813-178 has a mass smaller than that of the two detected WR stars.
Its mass is consistent with the  mass of the RSG,
and would require a mass at the TO of about 20 Msun, and an age of 5-7 Myr.   
Further observations are recommended to explore the degree of coevality in  Cl1813-178.
Some non-coevality would easily explain the discrepant masses.
A population with an  age of $4.0-4.5$ Myr, and a spread in age of
1 Myr, could explain the observed range of stellar masses.

The stellar mass at the  TO for a population with an age of $4.0-4.5$ Myr is likely between 25-35 \Msun. 
This is consistent with the  observations of dwarfs and giants.
Three  OB dwarfs with \Ks $< 10.0$ mag were detected with  O9-B3 spectral types, while the giants with 
O7-B3 types have \Ks=9.3-9.5 mag. An O7V star ($\sim 32$ \Msun) at an average extinction of  \Aks=0.8 mag,   
and a  distance of 4.8 kpc  is expected to have a \Ks=$\sim 10.4$ mag \citep{martins06}, while an O9V 
($\sim 22$ \Msun)  star has a \Ks=$10.9$ mag. 

A further spectroscopic survey  of fainter stars is needed to sample the TO region.
Moreover, high-resolution spectra  would allow for a more precise spectral classification. 
By narrowing the errors shown in Fig.\ \ref{fig.lum}, and increasing the sample, we will be able to 
better constrain the age, and verify the degree of coevality.

\section{The cluster mass}

By  adding the masses of the spectroscopically identified massive stars, we estimated a minimum 
cluster mass of 990 \Msun. We considered all 34 stars with $J-$\Ks\ between 1 and 3 mag, and \Ks $<
10$ mag. By assuming that they all have masses greater than 35 \Msun, and using a Salpeter mass
function integrated from 0.8 to 120 \Msun, we estimated a cluster mass of $13000\pm3000$ \Msun,
where the error is  the Poisson error of the number of massive stars. By assuming that these 34 stars
have  masses greater than 25 \Msun, we obtained a cluster mass of $8700\pm2000$ \Msun.  The two
calculations take into account the  uncertainties of the mass at the TO (25-35 \Msun), 
predicted for a coeval population with an age of 4-4.5  Myr. 

The Cl 1813-178 cluster is, therefore, a new addition to the list of 13 known young massive clusters
($\geq 10^4$ \Msun) in the Milky Way \citep{messineo09a,negueruela10}.

\section{Cluster surroundings}

\begin{figure*}[!]
\begin{centering}
\resizebox{1.0\hsize}{!}{\includegraphics[angle=0]{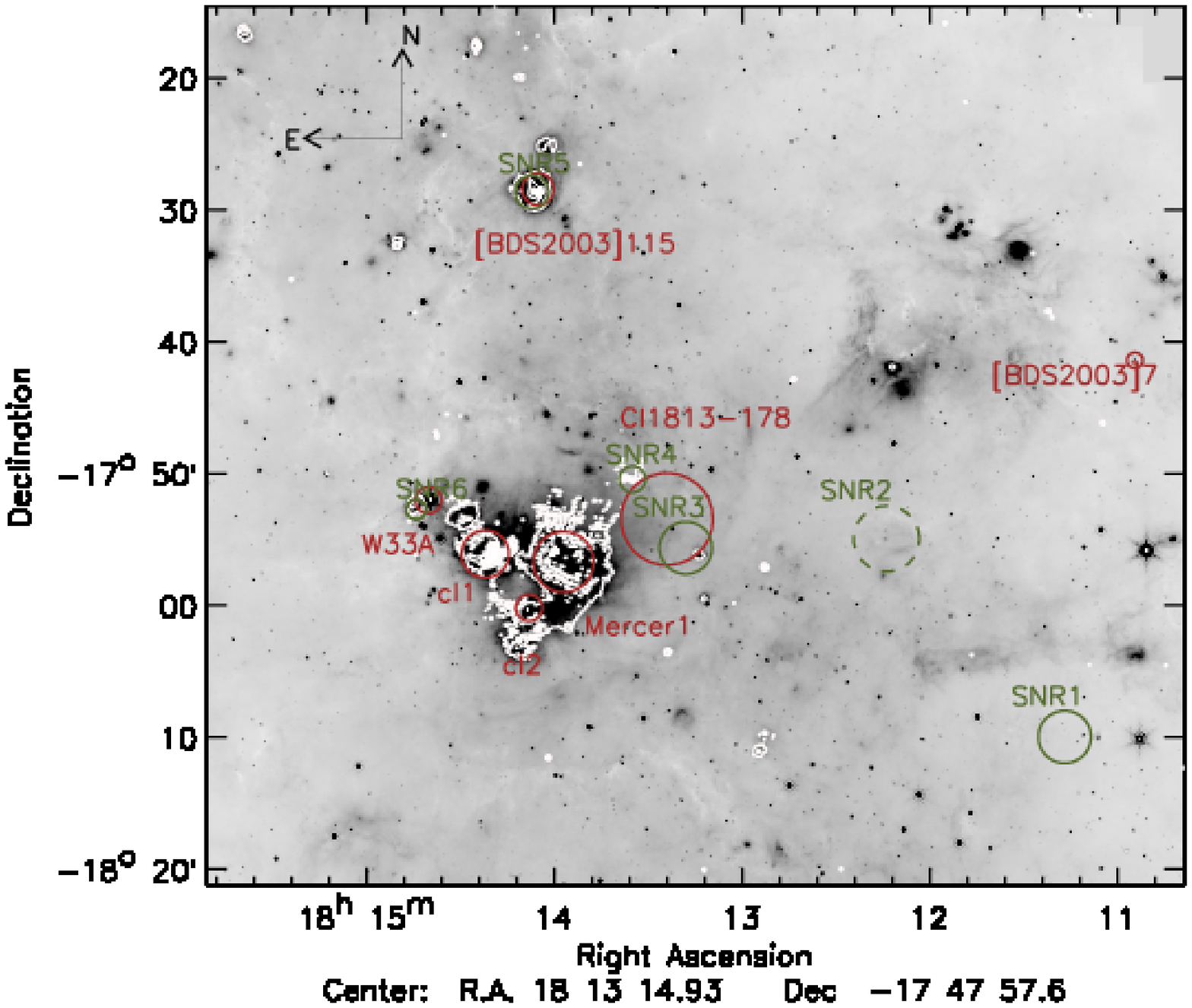}}
\end{centering}
\caption{\label{mipsigal:w33} 24 \um\ emission from MIPSGAL  \citep{carey09} of the whole W33 complex.
The white contours show  20 cm radio continuum emission from MAGPIS \citep{white05}.
Contour levels are 0.003, 0.009 and 0.020 mJy beam$^{-1}$; the beam size is 
6.2\arcsec $\times$ 5.4\arcsec, FWHM, and the position angle of the major axis is along 
the north-south direction.
Circles and labels indicate the location of the stellar clusters and candidate SNRs.}
\end{figure*}

A  24 \um\ image from MIPSGAL, the survey of the inner Galactic plane using the Multiband Infrared 
Photometer for Spitzer aboard the 
Spitzer Space Telescope \citep{carey09}, of the whole W33  region is shown
in  Fig.\ \ref{mipsigal:w33}, together with contours of radio continuum emission at 20 cm  \citep{white05}. 
The  complex  extends over an area of roughly 25\arcmin$\times$20\arcmin\ \citep{bieging78}.
Radio observations show that W33 is made up of a number of discrete sources. Some of these
radio sources have been classified as candidate SNRs  by \citet{brogan06} 
and \citet{helfand06} on the basis of morphology and spectral indexes, with MAGPIS data
\citep{white05}. See Table \ref{table.snr}.

The Cl 1813-178 cluster appears located on the edge of the W33 complex.  
The radial velocity of the K2I star in the Cl 1813-178 cluster is of 62$\pm$4  \kms\ (PaperI), 
and well agrees  with the high-velocity gas component of W33. The spatial coincidence of the
Cl 1813-178 cluster with the  SNR G12.82-0.02 and G12.72-0.00 has  already been reported in PaperI,
and support the association  of the cluster with the complex.
The filamentary shape of the Cl 1813-178 stellar cluster and its location on the edge
of W33 suggests a secondary episode of star formation, perhaps triggered by 
an expanding shell.

We searched for other possible candidate clusters and associations with SNRs in the direction of the W33 complex.
A number of candidate stellar clusters have been identified in GLIMPSE and 2MASS images 
in the direction of the W33 region by \citet{mercer05} and \citet{bica03}, which we list 
in Table \ref{table.clusters}. 
The  W33 MYSO \citep{davies10}  is projected into SNR6 (G13.1875+0.0389) \citep{helfand06}. 
The spatial coincidence of  the SNR and the MYSO suggests a physical 
association.  The  W33A MYSO could be an episode of triggered star formation induced by a supernova explosion. 
Mercer1 candidate cluster is the object number \#1 in the list of \citet{mercer05}.
It was identified as a stellar overdensity  in the GLIMPSE catalog with an 
automatic algorithm. It is located at the center of the molecular complex, and
it appears as a spread overdensity. 
Two other candidates are reported in literature in the surrounding of the W33 complex,
but without a clear connection with the W33 complex. The BDS2003-115 candidate cluster is  about 20\arcmin\ 
North of the main W33 complex and is associated  with  SNR5 (G12.83-0.02) \citep{helfand06}. 
The BDS2003-7 candidate cluster appears as a small group of stars  without associated 
radio emission \citep{bica03}.

\begin{figure*}[!]
\begin{centering}
\resizebox{0.32\hsize}{!}{\includegraphics[angle=0]{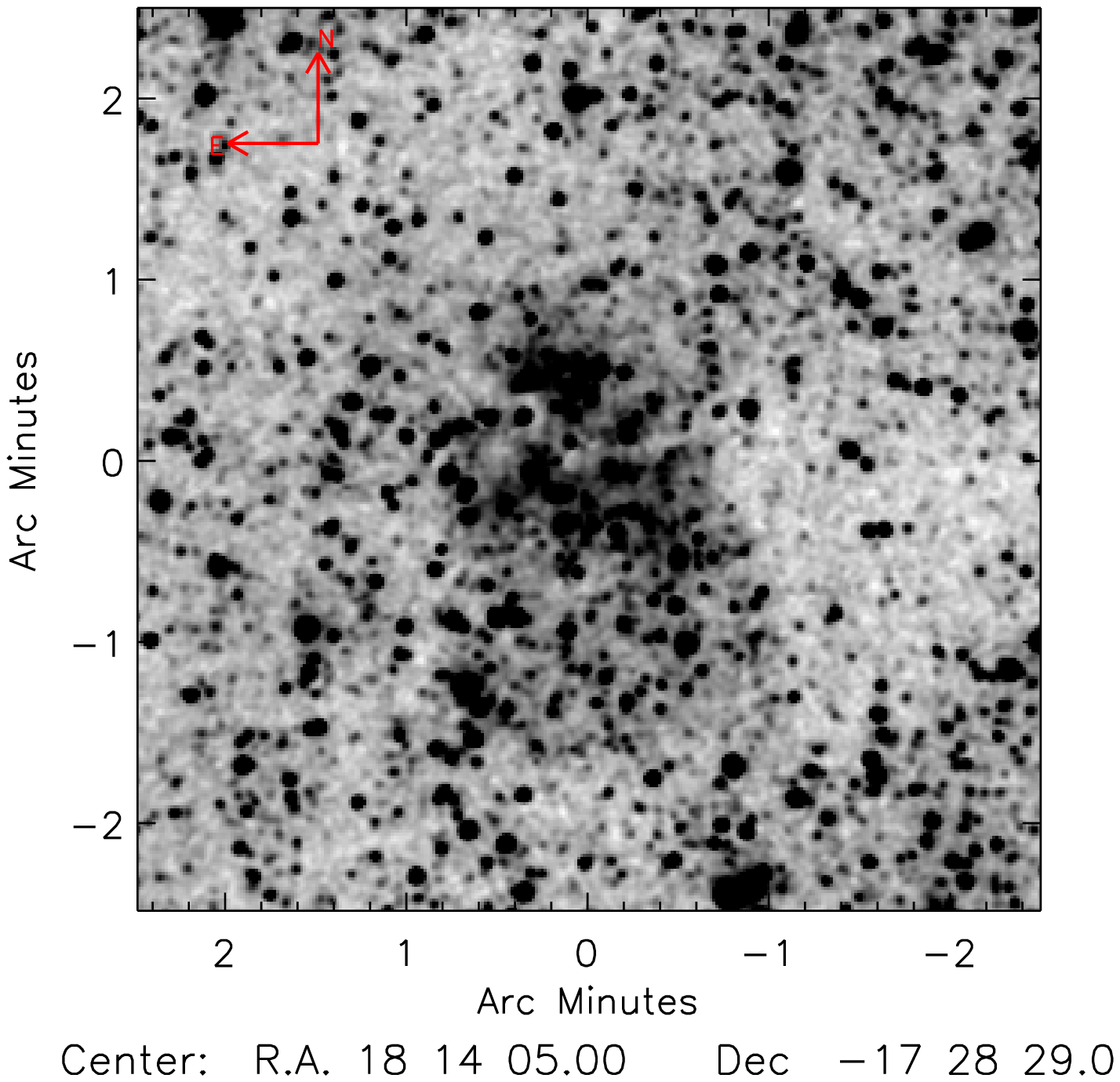}}
\resizebox{0.32\hsize}{!}{\includegraphics[angle=0]{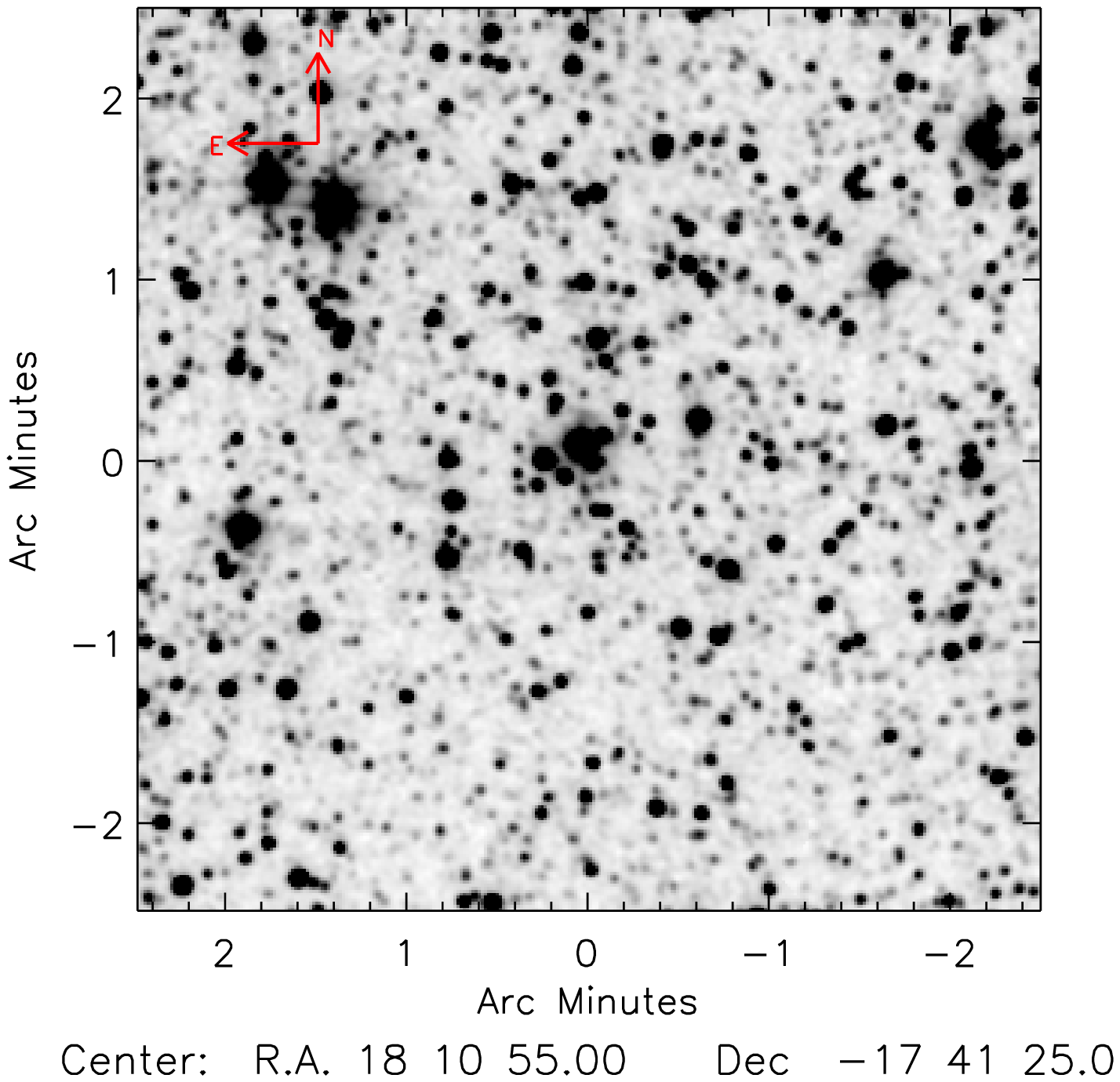}}
\resizebox{0.32\hsize}{!}{\includegraphics[angle=0]{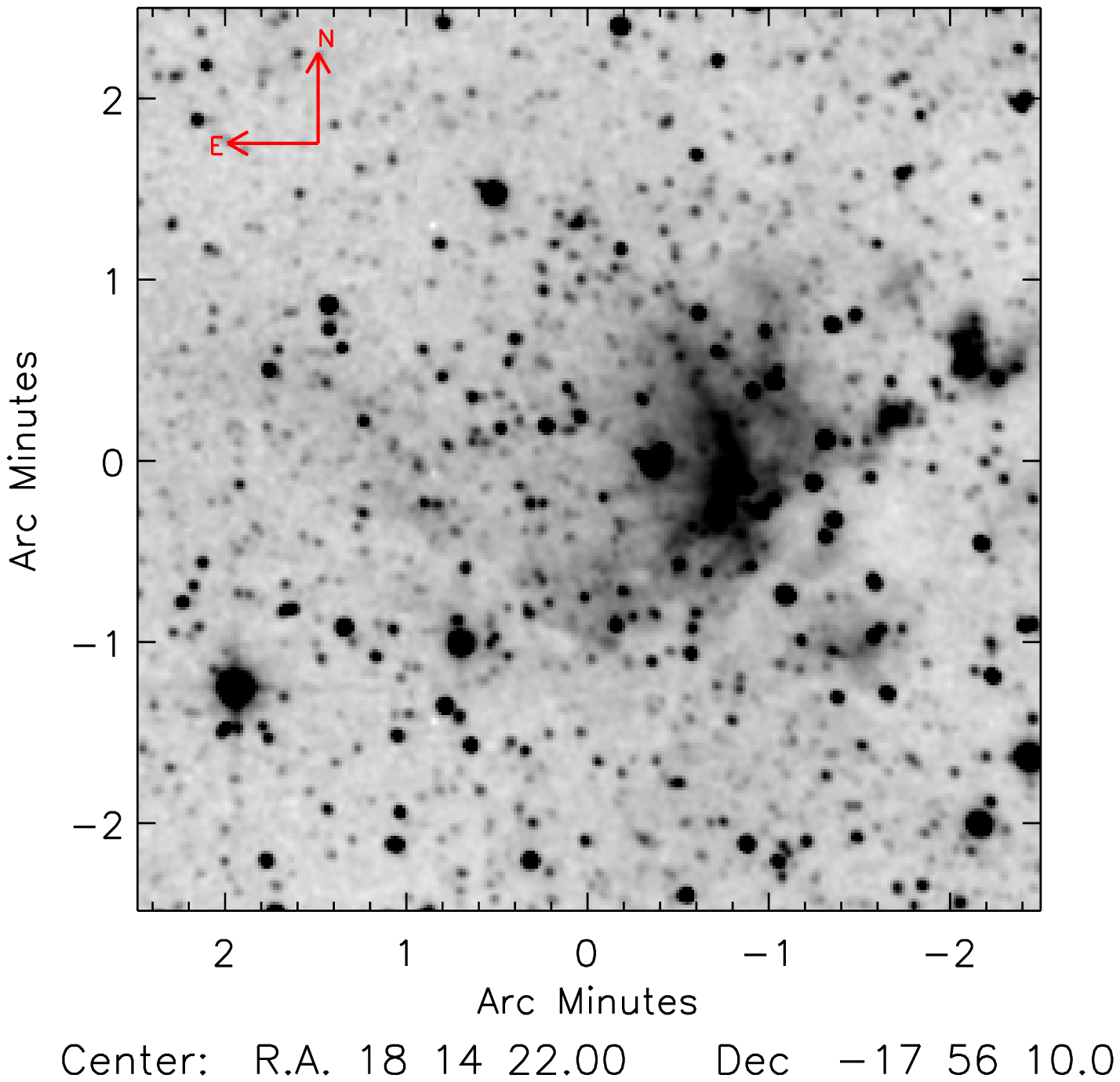}}
\resizebox{0.32\hsize}{!}{\includegraphics[angle=0]{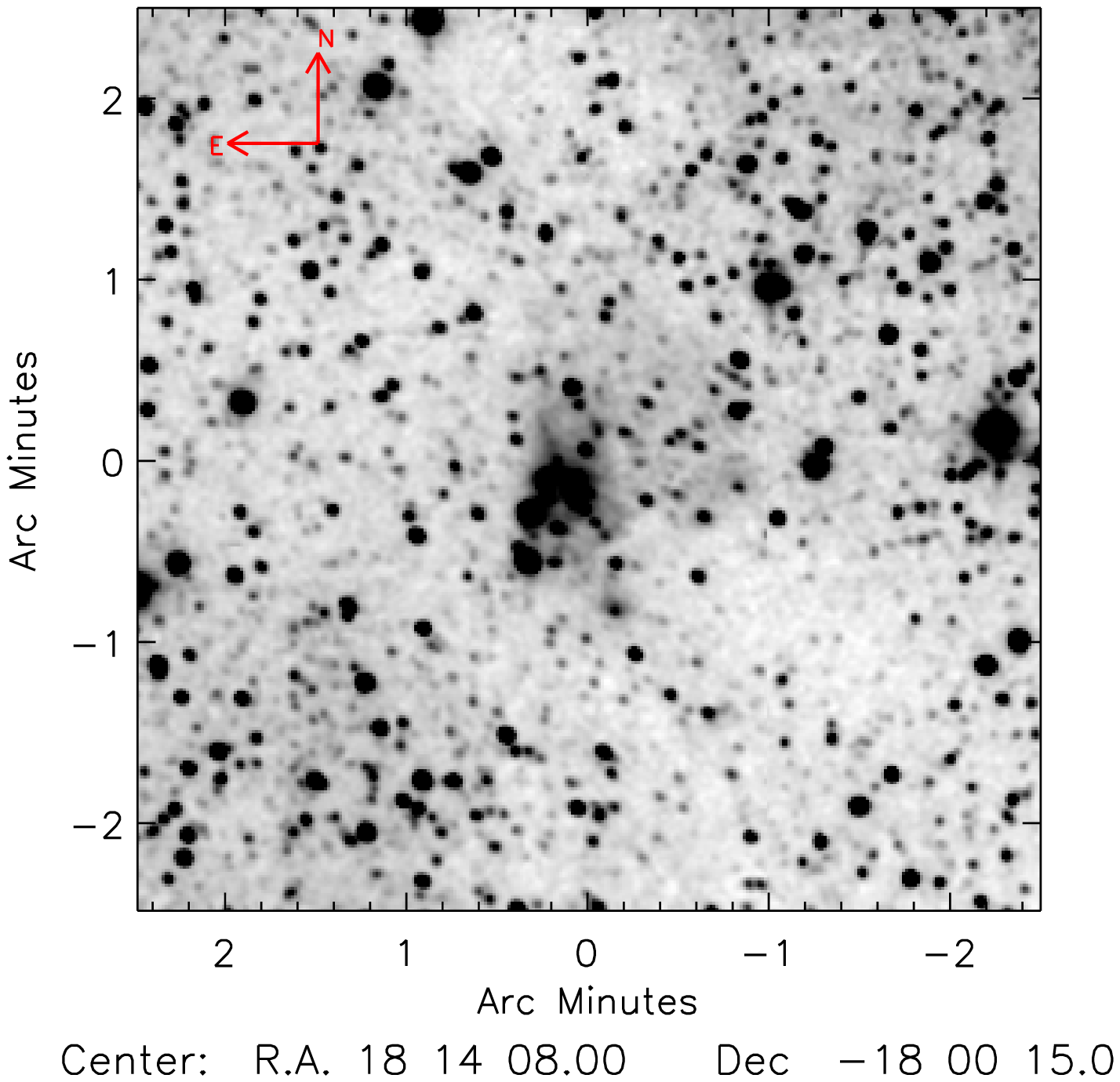}}
\resizebox{0.32\hsize}{!}{\includegraphics[angle=0]{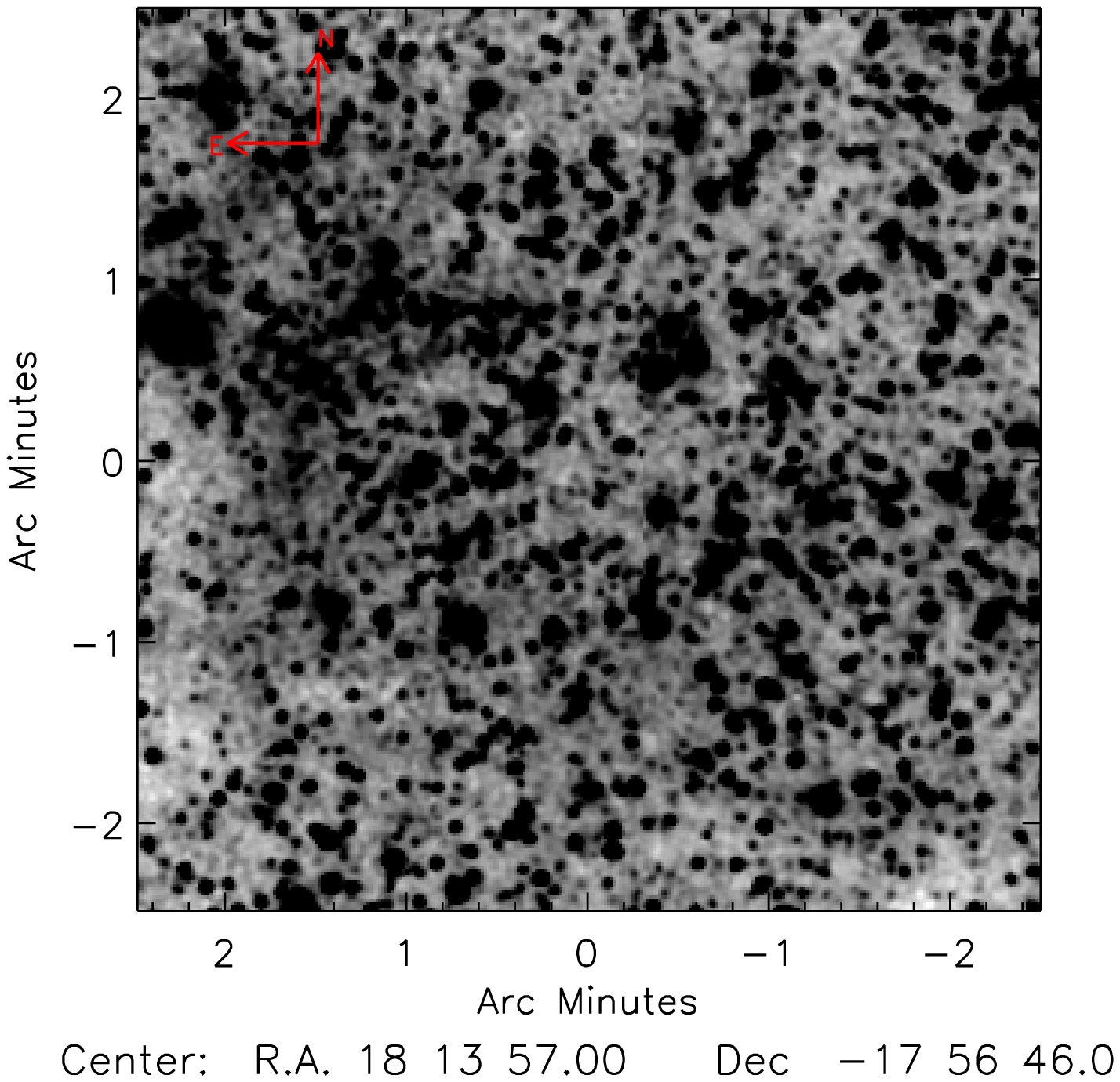}}
\end{centering}
\caption{\label{fig.cand1} 
2MASS \Ks\ images  of the  BDS2003-115 cluster (top-left),
the  BDS2003-7 cluster (top-middle), the  cl1 cluster (top-right),
the cl2 cluster (bottom-left), and the Mercer1 cluster (bottom-middle). }
\end{figure*}

\begin{figure*}[!]
\begin{centering}
\resizebox{0.32\hsize}{!}{\includegraphics[angle=0]{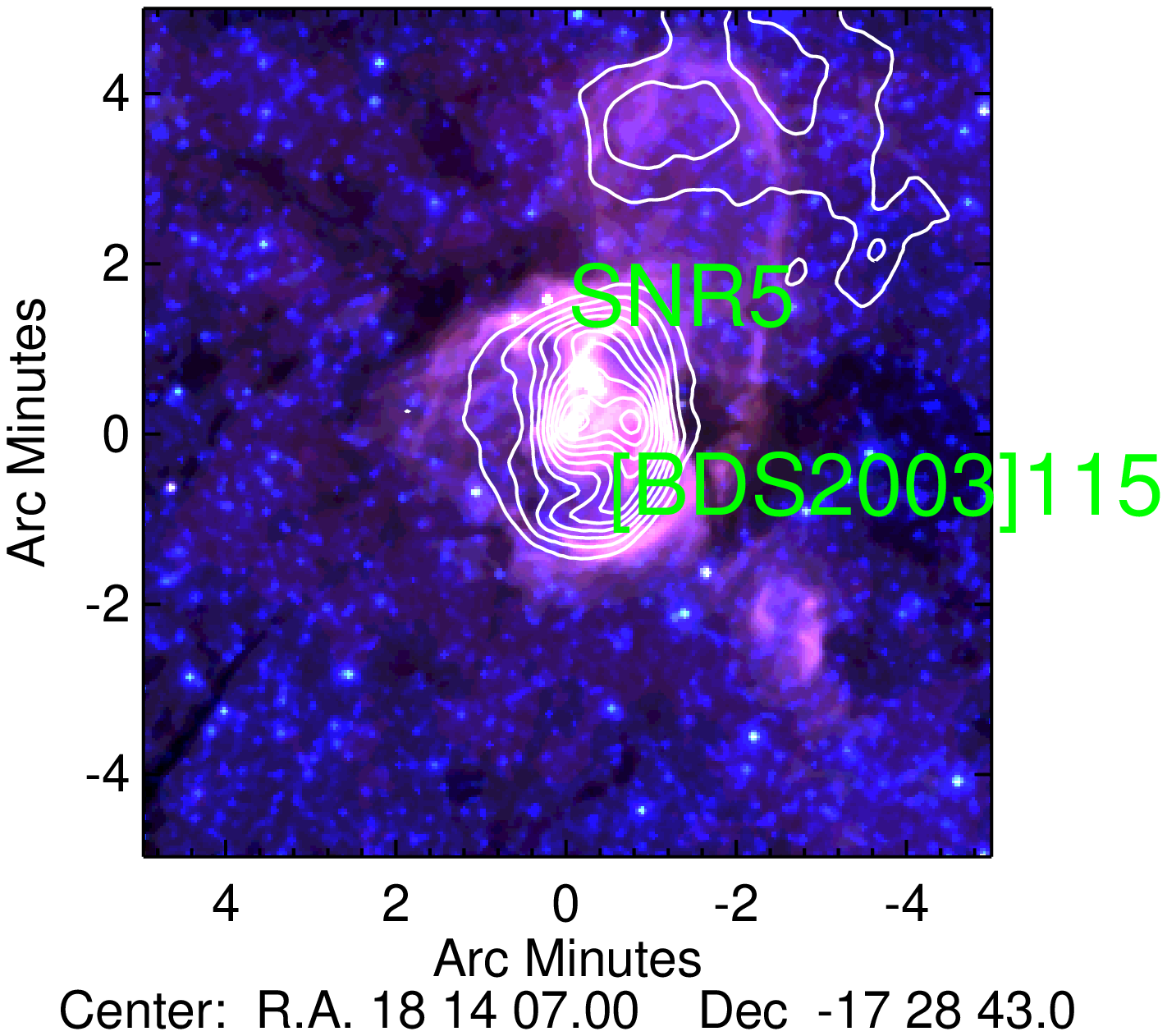}}  
\resizebox{0.32\hsize}{!}{\includegraphics[angle=0]{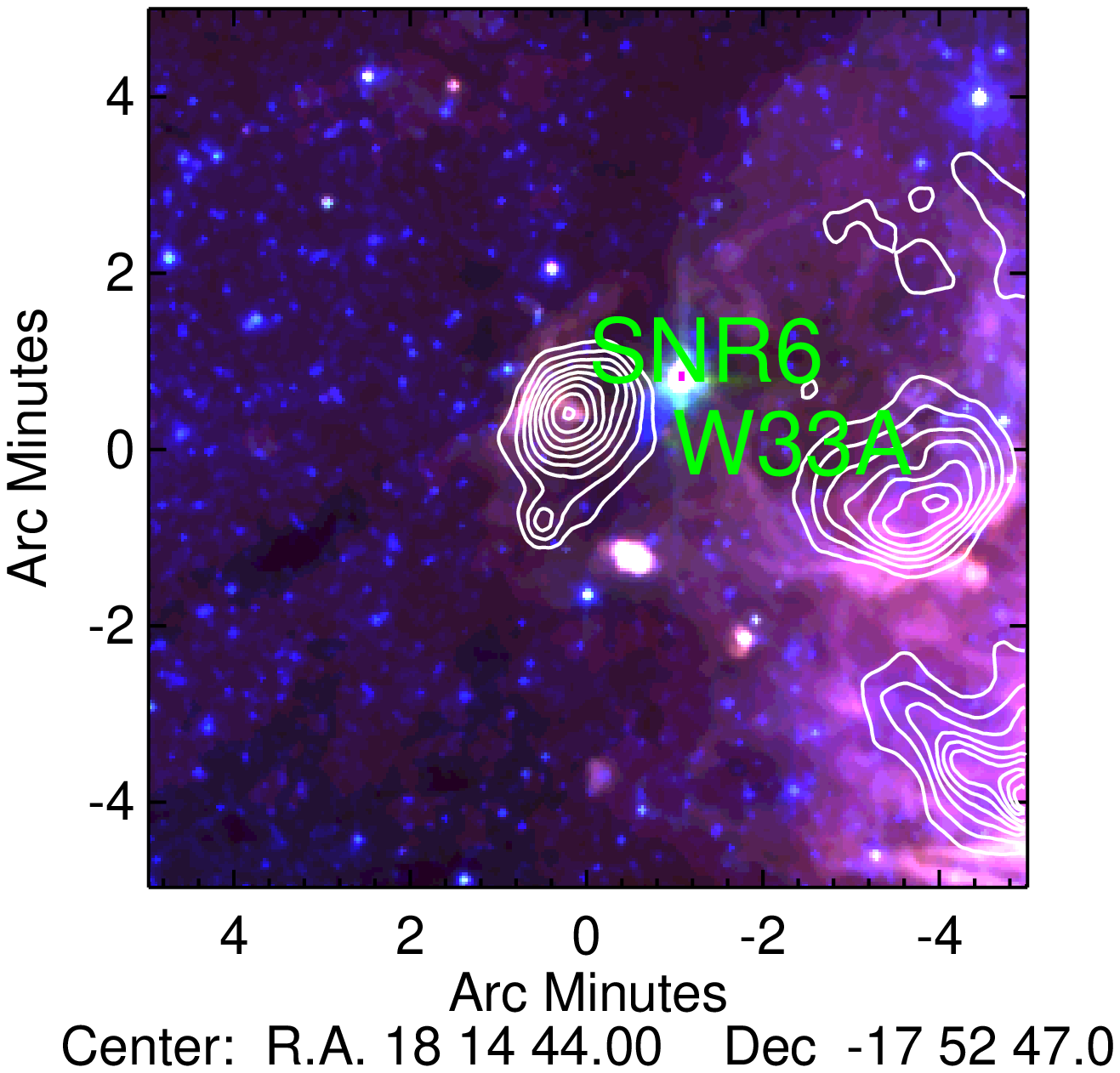}}  
\resizebox{0.32\hsize}{!}{\includegraphics[angle=0]{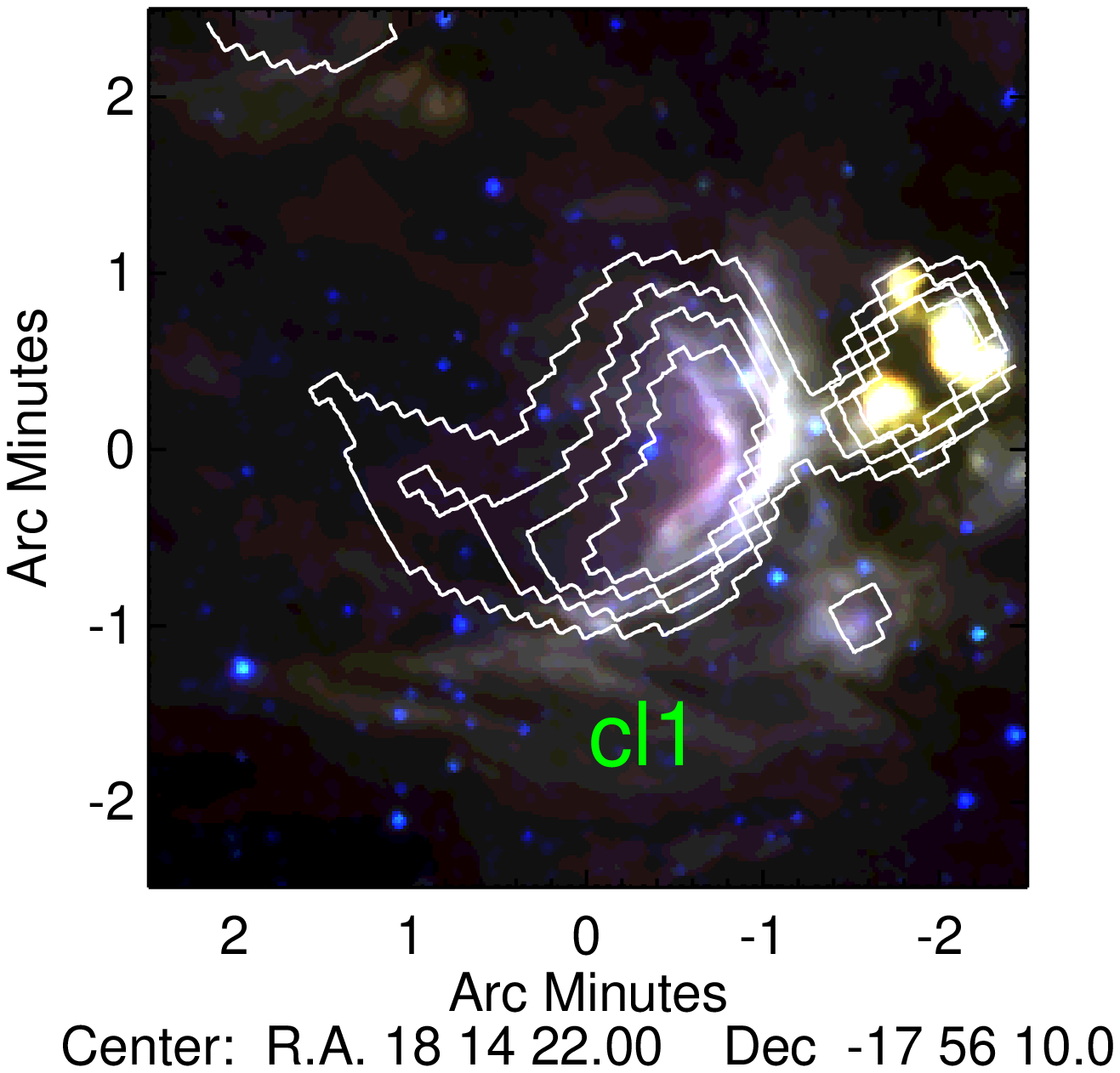}}  
\resizebox{0.32\hsize}{!}{\includegraphics[angle=0]{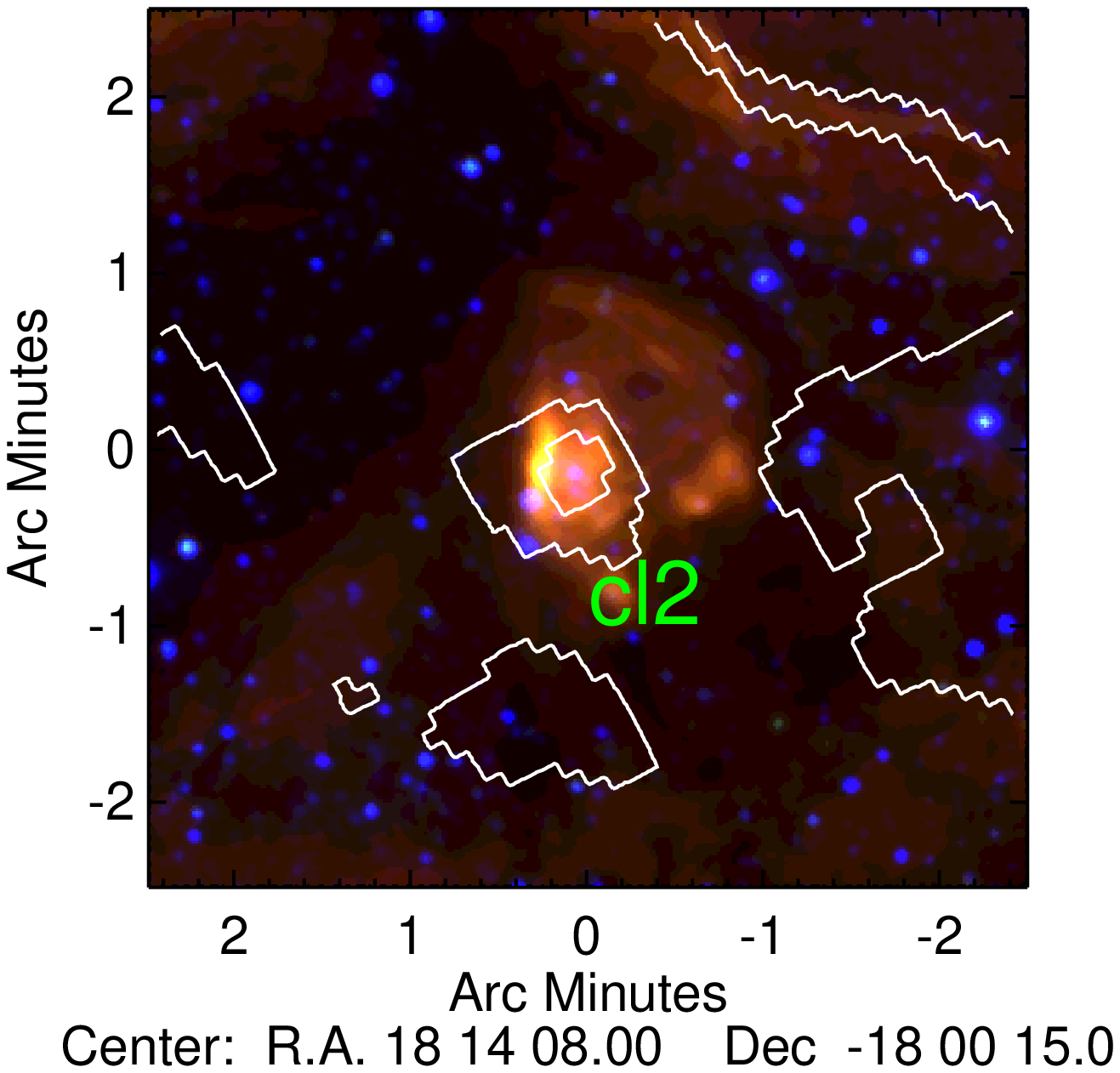}}  
\resizebox{0.32\hsize}{!}{\includegraphics[angle=0]{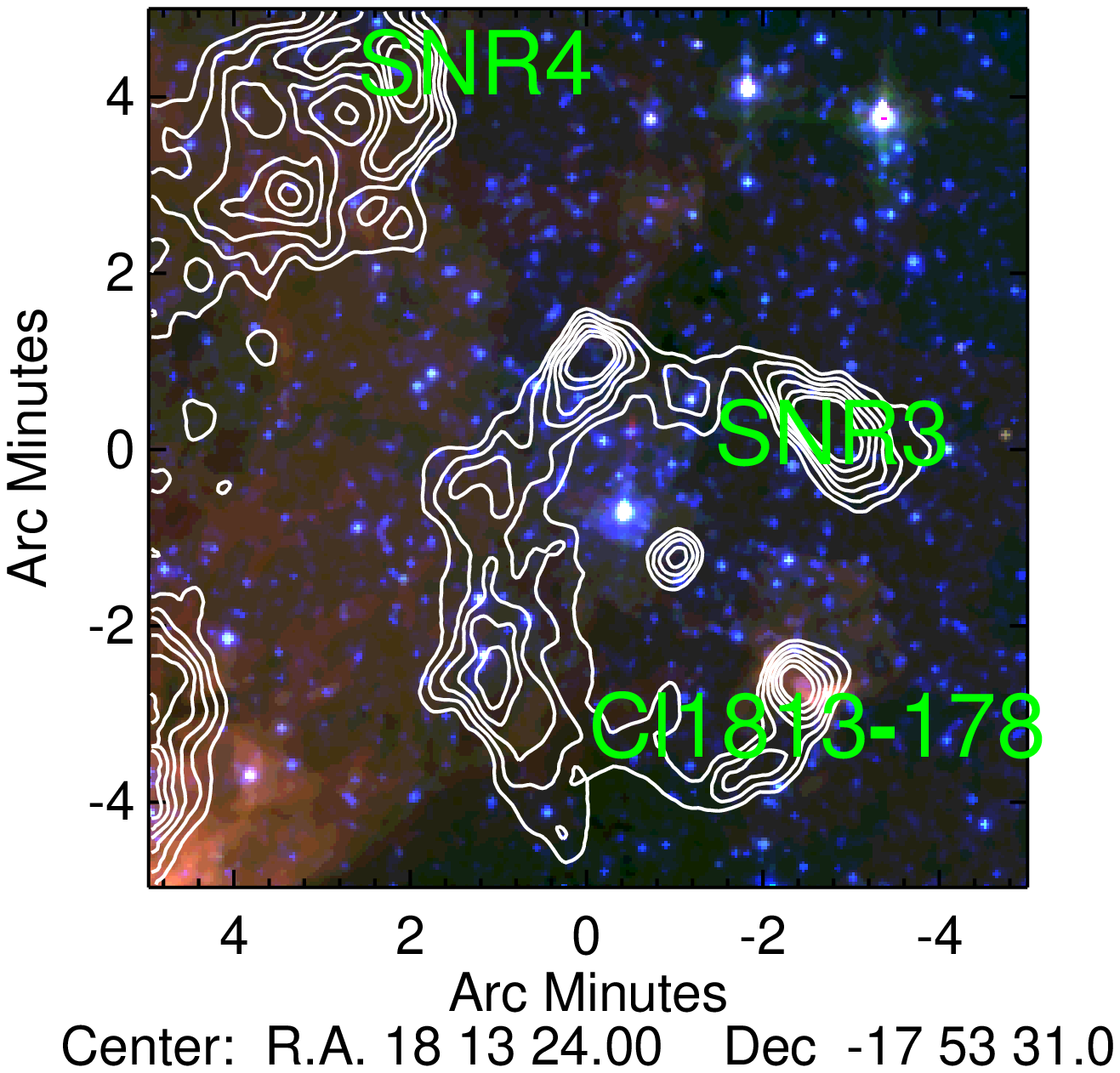}}  
\end{centering}
\caption{\label{fig.cand2} 
Composite images of SNR5, SNR6, cl1, cl2, and Cl 1813-178. 
In blue the GLIMPSE 3.6 \um, in green the 5.8\um, and in red the 8.0\um. Over-plotted 
are contours of  90 cm radio continuum emission detected by MAGPIS \citep{white05}. }
\end{figure*}

We visually inspected the 2MASS images, and located two other  clumps of stars
(cl1 and cl2).   The cl1 candidate appears as a group of point sources on bright
nebular emission in the 2MASS \Ks\  image. Inspection of the GLIMPSE and MAGPIS
images reveals the presence of an HII region,  suggesting the presence of
massive stars (see Table \ref{table.clusters} and Figs.  \ref{fig.cand1} and
\ref{fig.cand2}). The cl2 candidate is another small concentration of bright
stars (\Ks=8-10 mag) in  another HII region. Nothing
is reported in previous literature about both, cl1 and cl2, clumps.
In addition, we searched for  stellar over-densities in the direction 
of the W33 complex using both 2MASS and GLIMPSE star-counts.
Detections are hampered by strong variations of the background level, 
which are due to variations of interstellar extinction and nebular emission.
A spectroscopic and photometric follow-up study of these  regions
with SINFONI and UKIDSS data is ongoing, in order to
confirm the presence of massive stars.

Near-infrared spectroscopic follow-up observations are needed to characterize these sources, 
and to confirm their association with the W33 complex. However, the 
associations of Cl 1813-178, cl1, and cl2 with HII regions and/or SNRs 
suggest that  these are other condensation of massive stars in  
the W33 complex.

\begin{deluxetable}{lrrrr}
\tablewidth{0pt}
\tablecaption{\label{table.clusters} List of candidate clusters in the direction of the W33 complex.}
\tablehead{
\colhead{ID}& 
\colhead{RA}& 
\colhead{DEC}& 
\colhead{Radius(\arcmin)} &
\colhead{References} 
}
\startdata
Cl 1813-178      & 18 13 24  &-17 53 31 &3.5 & PaperI\\
Mercer1          & 18 13 57  &-17 56 46 &2.3 &\citep{mercer05}\\
BDS2003-115      & 18 14 05  &-17 28 29 &1.2 &\citep{bica03}\\
BDS2003-7        & 18 10 55  &-17 41 25 &0.6 &\citep{bica03}\\
cl1              & 18 14 22  &-17 56 10 &1.8 & Present work\\
cl2              & 18 14 08  &-18 00 15 &1.0 & Present work\\
\enddata
\end{deluxetable}

\begin{deluxetable}{llrrrll}
\tablecaption{\label{table.snr} List of candidate SNRs}
\tablehead{
\colhead{ID}& 
\colhead{Name}& 
\colhead{RA}& 
\colhead{DEC} &
\colhead{Radius(\arcmin)} &
\colhead{References} 
}
\startdata
SNR1 &G12.26+0.30     &  18:11:17     & -18:10:00    & 4   & \citep{brogan06,helfand06}\\            
SNR2 &G12.58+0.22 ?   &  18:12:14     & -17:55:00    & 5   & \citep{brogan06}\\                      
SNR3 &G12.72-0.00     &  18:13:18     & -17:55:42    & 4   & \citep{brogan06,helfand06}\\ 
SNR4 &G12.83-0.02     &  18:13:35     & -17:50:30    & 2   & \citep{brogan06,helfand06}\\ 
SNR5 &G13.1875+0.0389 &  18:14:07     & -17:28:43    & 2.5 & \citep{helfand06}\\
SNR6 &G12.9139-0.2806 &  18:14:44     & -17:52:47    & 1.5 & \citep{helfand06}\\
\enddata
\end{deluxetable}

\section{Summary}
A near-infrared spectroscopic survey of the brightest stars in the direction of 
the Cl 1813-178 cluster is presented. Among the 61  observed stars,
25 massive stars were detected. Two WR stars of type WN7,   a cLBV, 
and 21 OB stars were  identified.
Among the OB stars, a O8-O9If star and a O6-O7If star were discovered.
Eight of these evolved stars also have X--ray emission, as detected
by the Chandra and XMM satellites. The hardness of the X--ray emission from the
two WN7 stars strongly suggests binary systems.

A  spectro-photometric analysis of the OB stars reveals 14 supergiants, 4 giants, and 3 dwarfs.
From the giants, dwarfs and WRs, we derived average spectrophotometric distances of  $3.8\pm1.0$ kpc,
and $2.9\pm0.8$ kpc, and $4.2\pm1.6$ kpc. The distances  from giants and WRs
is in agreement with the kinematic distance.
The distance estimates from dwarfs  are only consistent with the kinematic distance if the 
dwarfs are late O stars.

The mixture of evolved massive stars is reminiscent of other Galactic young massive clusters, 
such as Westerlund 1,  Quintuplet, Galactic center, and  Cl 1806-20.
We estimated stellar luminosities, therefore, masses by comparing the luminosities
with evolutionary tracks from the Geneva group.  
By assuming a Salpeter mass function, we obtained a cluster mass of $1.0\pm0.2 \times 10^4$ \Msun.
A likely cluster age of 4-4.5 Myr is derived, however, a spread in age of about 1 Myr cannot be excluded. 
In order to better constrain the degree of coevality, further spectroscopic observations
are required.

The Cl 1813-178 cluster is located on the Western edge of the W33 complex.
We have located several other candidate stellar clusters that could belong to the same complex.

\acknowledgments {
This work was partially funded by the ERC Advanced Investigator Grant GLOSTAR (247078). 
The material in this work is partly supported by NASA under
award NNG 05-GC37G, through the Long--Term Space Astrophysics
program. Part of this research was performed in the Rochester
Imaging Detector Laboratory with support from a NYSTAR Faculty
Development Program grant.
We also acknowledge support from the US National Science Foundation under
grant  AST-0709479 (R.M.R.).
RPK acknowledges support by the Alexander-von-Humboldt Foundation.
F.N. acknowledges financial support from the Spanish Ministerio 
de Ciencia e Innovacion under the project AYA2008-06166-
C03-02.
This publication makes use of data products from the Two
Micron All Sky Survey, which is a joint project of the University of
Massachusetts and the Infrared Processing and Analysis
Center/California Institute of Technology, funded by the National
Aeronautics and Space Administration and the NSF. This research has
made use of Spitzer's GLIMPSE survey data, the Simbad and Vizier
database. We thank  the staff of the Joint Astronomy Centre  for their
great support during the UKIRT run.
We thank the staff astronomers of the European Southern Observatory
for their excellent support during the SofI observations.
We thank Dr. Thomas Dame for kindly providing us with a CO
map of the W33 complex. A special thank to Dr. Andreas Brunthaler
for discussion on parallactic distances to the W33 complex using methanol masers.
We acknowledge helpful comments and suggestions
by an anonymous referee.}

\appendix{
\section{Late-type stars}

The  EW(CO)s can be used for classification  in subclasses of   late-type stars.
There is a linear correlation between stellar effective temperatures and EW(CO)s of the CO bands.
Furthermore, since giants and supergiants follow two different relations, information on the luminosity class can
also be obtained \citep[see e.g.][]{figer06,davies07}.

Among the observed 60 stars, 37 are found to be late-type stars.
For  36 of them, spectra were taken with the K-long grism, covering the region of the 
CO band-head at 2.29 \um.  CO bands at 
2.29 \um\ in absorption are detected in all spectra. 
Star \#55 was observed only  with the K-short grism;
however, the presence of  Na lines (at 2.2075 and 2.2077 \um)  and Mg I at 2.11 \um\ suggests a
spectral type later than G0.

The equivalent widths of 32 of these late-type stars are compatible with that of giant
stars with spectral type from K0III to M7III (Table \ref{table.nonmembers}).
For star \#58 we do not report any spectral type, since its spectrum  has a poor signal to noise.

Star \#32, \#38, and  \#39 have EW(CO)s  typical of RSG stars, with spectral types
M2.5, M3.5 and M1, respectively.  However, their photometric properties
indicate that they are unrelated to the stellar cluster.
Star \#1 is consistent with  being a K2-K5I  cluster member 
(see PaperI).

Young massive clusters with ages from 4 to 30 Myrs may contain yellow supergiants (YSG) stars, e.g. the
Westerlund 1 \citep{clark05} and RSGC1 clusters \citep{figer06}.  YSG are rare F or G-type supergiants in
transition towards  the RSG phase or, back from the RSG locus,  evolving blue-ward. In a coeval population,
YSGs or RSGs  are expected to be  brighter in \Ks\ than early-type massive stars. In the Cl 1813-178
cluster, all  detected stars with CO band-head at 2.29 \um\ appear fainter than the brightest OB stars, and
the gap between the brightest early-type star (\Ks=6.75 mag) and the K2I star (\Ks=3.79 mag) is devoid of
stars. Furthermore, the  CO band-heads  of the five late-type stars with \Ks\ between 7 and 8 mag indicate
late M giants (M2-M5), or late K supergiants (K2-K4).  Their  EW(CO)s are not consistent with being F or G
supergiants.  

{\small
\begin{deluxetable}{lrrrrrrrrrrrrllr}
\rotate
\tablewidth{0pt}
\tabletypesize{\tiny}
\tablecaption{\label{table.nonmembers} List of field stars spectroscopically observed.}
\tablehead{
\colhead{ID}& 
\colhead{Ra}& 
\colhead{Dec}& 
\colhead{B} & 
\colhead{V} & 
\colhead{R} &
\colhead{J} & 
\colhead{H} & 
\colhead{K$_{\rm s}$} &
\colhead{[3.6]} & 
\colhead{[4.5]} &
\colhead{[5.8]} & 
\colhead{[8.0]} & 
\colhead{ tel.} &
\colhead{Sp} 
}
\startdata
10  & 18 13 47.56  & -17 57 01.43  &  15.32  &\nodata  &  13.64  &  10.76  &   9.89  &   9.60  &   9.31  &   9.34  &   9.22  &   9.20  &      ukirt &   K0.5III  \\
27  & 18 13 14.39  & -17 54 47.87  &\nodata  &\nodata  &  17.48  &   9.50  &   7.84  &   7.08  &   6.90  &   6.81  &   6.55  &   6.54  & ukirt-keck &   M4.5III  \\
28  & 18 13 35.03  & -17 54 26.11  &  12.85  &  12.70  &\nodata  &   8.53  &   7.53  &   7.21  &   7.16  &   7.22  &   7.00  &   7.03  &       keck &     M4III  \\
29  & 18 13 29.14  & -17 52 27.18  &  16.19  &  13.16  &  13.13  &   9.00  &   7.94  &   7.50  &   7.26  &   7.51  &   7.26  &   7.28  &       keck &     M3III  \\
30  & 18 13 26.87  & -17 55 27.50  &\nodata  &\nodata  &\nodata  &  10.35  &   8.40  &   7.52  &   7.07  &   7.12  &   6.90  &   6.87  &       keck &     M5III  \\
31  & 18 13 16.80  & -17 55 40.02  &  21.42  &  16.78  &  14.36  &   9.79  &   8.47  &   7.93  &   7.63  &   7.72  &   7.58  &   7.55  &       keck &   M2.5III  \\
32  & 18 13 16.35  & -17 50 29.89  &\nodata  &\nodata  &\nodata  &  13.53  &   9.85  &   8.08  &   6.82  &   6.80  &   6.41  &   6.47  &       keck &   M2.5I    \\
33  & 18 13 25.93  & -17 57 39.84  &  16.25  &  16.37  &\nodata  &  11.61  &   9.27  &   8.24  &   7.55  &   7.66  &   7.39  &   7.41  &       keck &   M5.5III  \\
34  & 18 13 28.59  & -17 50 30.78  &  12.42  &  11.65  &  11.15  &   9.10  &   8.51  &   8.40  &   8.34  &   8.40  &   8.31  &   8.27  &       keck &   K0.5III  \\
35  & 18 13 34.10  & -17 51 28.59  &\nodata  &\nodata  &\nodata  &  11.55  &   9.40  &   8.49  &   7.78  &   7.92  &   7.64  &   7.66  &       keck &     M2III  \\
36  & 18 13 31.44  & -17 52 41.01  &\nodata  &\nodata  &\nodata  &  14.13  &  10.40  &   8.50  &   7.00  &   6.83  &   6.35  &   6.43  & ukirt-keck &     M7III  \\
37  & 18 13 21.37  & -17 58 04.63  &  15.18  &  13.87  &  14.64  &   9.79  &   8.81  &   8.51  &   8.36  &   8.46  &   8.28  &   8.19  &       keck &   K3.5III  \\
38  & 18 13 28.96  & -17 55 52.85  &\nodata  &\nodata  &\nodata  &  17.08  &  11.22  &   8.54  &   6.89  &   6.19  &   5.63  &   5.41  &       keck &   M3.5I    \\
39  & 18 13 29.36  & -17 51 47.65  &\nodata  &\nodata  &\nodata  &  13.83  &  10.34  &   8.61  &   7.50  &   7.49  &   7.08  &   7.18  &      ukirt &   M1I      \\
40  & 18 13 14.67  & -17 51 32.41  &  16.43  &  14.40  &  13.69  &  10.08  &   9.06  &   8.69  &   8.45  &   8.53  &   8.39  &   8.37  &      ukirt &   K5.5III  \\
41  & 18 13 12.16  & -17 51 58.47  &\nodata  &\nodata  &\nodata  &  11.60  &   9.63  &   8.78  &   8.17  &   8.24  &   8.04  &   8.03  &      ukirt &     K5III  \\
42  & 18 13 19.97  & -17 53 42.39  &\nodata  &\nodata  &\nodata  &  13.94  &  10.46  &   8.80  &   7.68  &   7.68  &   7.33  &   7.44  &      ukirt &     M6III  \\
43  & 18 13 21.98  & -17 53 20.70  &\nodata  &\nodata  &\nodata  &  13.25  &  10.17  &   8.84  &   7.83  &   7.82  &   7.46  &   7.35  &      ukirt &     M3III  \\
44  & 18 13 14.87  & -17 57 21.60  &  18.60  &\nodata  &  18.54  &  14.39  &  10.65  &   8.86  &   7.58  &   7.56  &   7.18  &   7.27  &      ukirt &     M7III  \\
45  & 18 13 19.55  & -17 55 01.00  &  17.16  &  17.08  &\nodata  &  13.19  &  10.30  &   8.96  &   8.05  &   8.05  &   7.76  &   7.82  &      ukirt &   M2.5III  \\
46  & 18 13 25.70  & -17 52 01.02  &\nodata  &\nodata  &\nodata  &  13.60  &  10.45  &   8.99  &   7.97  &   7.95  &   7.63  &   7.66  &      ukirt &   M6.5III  \\
47  & 18 13 12.25  & -17 51 26.83  &\nodata  &\nodata  &\nodata  &  13.14  &  10.36  &   9.06  &   8.22  &   8.31  &   7.92  &   8.03  &      ukirt &     M6III  \\
48  & 18 13 17.58  & -17 54 52.09  &\nodata  &\nodata  &\nodata  &  14.00  &  10.63  &   9.07  &   7.91  &   7.99  &   7.66  &   7.68  &      ukirt &     M7III  \\
49  & 18 13 25.18  & -17 49 35.98  &  13.82  &  13.12  &  12.08  &  10.14  &   9.38  &   9.12  &   9.02  &   9.18  &   9.00  &   9.03  &      ukirt &     K3III  \\
50  & 18 13 10.77  & -17 54 47.58  &\nodata  &\nodata  &\nodata  &  11.51  &   9.83  &   9.13  &   8.74  &   8.77  &   8.56  &   8.62  &      ukirt &   K5.5III  \\
51  & 18 13 18.55  & -17 56 18.30  &\nodata  &\nodata  &  18.42  &  11.17  &   9.73  &   9.16  &   8.78  &   8.84  &   8.67  &   8.68  &      ukirt &     K4III  \\
52  & 18 13 35.17  & -17 50 44.06  &\nodata  &\nodata  &\nodata  &  13.90  &  10.68  &   9.19  &   8.15  &   8.10  &   7.78  &   7.92  &      ukirt &   M2.5III  \\
53  & 18 13 38.15  & -17 57 07.51  &\nodata  &\nodata  &\nodata  &  15.51  &  11.31  &   9.26  &   7.95  &   7.90  &   7.40  &   7.67  &      ukirt &   M6.5III  \\
54  & 18 13 35.19  & -17 57 33.25  &  12.71  &  12.33  &  11.85  &  10.00  &   9.46  &   9.30  &   9.28  &   9.31  &   9.19  &   9.21  &      ukirt &     K0III  \\
55  & 18 13 27.03  & -17 50 27.76  &  15.43  &  14.38  &  15.62  &  10.64  &   9.66  &   9.38  &   9.23  &   9.22  &   9.25  &   9.17  &      ukirt &   $>$G     \\  
56  & 18 13 18.70  & -17 51 14.93  &  20.67  &  15.81  &  16.69  &  10.90  &   9.76  &   9.40  &   9.21  &   9.25  &   9.19  &   9.24  &      ukirt &    K5.5III \\
57  & 18 13 17.23  & -17 55 21.81  &  15.93  &  13.67  &  15.02  &  10.61  &   9.83  &   9.56  &   9.36  &   9.45  &   9.33  &   9.23  &      ukirt &    K0.5III \\
58  & 18 13 12.99  & -17 54 02.32  &\nodata  &\nodata  &\nodata  &  14.55  &  11.42  &  10.02  &   9.06  &   9.01  &   8.71  &   8.75  &      ukirt &    $>G$   \\
59  & 18 13 20.01  & -17 53 50.13  &  15.83  &  14.87  &  15.93  &  11.54  &  10.73  &  10.52  &  10.32  &  10.42  &  10.34  &  10.15  &      ukirt &    K1.5III \\
60  & 18 13 20.65  & -17 53 50.28  &\nodata  &\nodata  &\nodata  &  16.06  &  12.65  &  10.58  &  10.04  &   9.97  &   9.70  &   9.70  &      ukirt &    M6.5III \\
61  & 18 13 26.31  & -17 53 57.05  &\nodata  &  17.34  &\nodata  &  14.99  &  13.30  &  11.77  &  10.73  &  10.62  &  10.35  &  10.48  &      ukirt &    K4.5III \\
\enddata		 																         
\tablecomments{For each star, number designations and coordinates
(J2000) are followed by magnitudes measured in different bands.
\J,\H, and \Ks\ measurements are from 2MASS, while the magnitudes at 3.6 \um,
4.5 \um, 5.8 \um, and 8 \um\ are from GLIMPSE.  $B$, $V$, and $R$
associations are taken from the astrometric catalog NOMAD. }
\end{deluxetable}
}
}

\section{Effective temperatures and bolometric corrections}

In order to estimate stellar luminosities, estimates of  effective temperatures and bolometric corrections 
as a function of spectral type need to be known.
Since an homogeneous  calibration extending from O stars down to early A stars
is missing, we summarize all the adopted values.  

For O-type stars, we used the  \Teff\ from \citet{martins05}. For early B supergiants, we adopted the \Teff\  
values given by \citet{crowther06b}, while,  for late B and A supergiants, those by \citet{humphreys84}. 
For B giants, we adopted the \Teff\ by \citet{humphreys84}.
For B and A dwarfs, we used \Teff\ estimated by \citet{humphreys84} and \citet{johnson66}.

For O-type stars, we used the bolometric corrections in K-band (\BCK) by \citet{martins06}.
For early B-type stars, we used those  provided by \citet{bibby08}.
We estimated the \BCK s of late B  supergiants and  giants by assuming bolometric corrections 
in the V-band (\BCV) from \citet{humphreys84}, and intrinsic $V-K$ colors
from \citet{koornneef83} and \citet{wegner94}.
For dwarf stars, a set of homogeneous \BCK\ was obtained by interpolating  an isochrone of 0.5 Myr 
and solar metallicity from \citet{lejeune01} at the assumed effective temperatures
\citep{humphreys84,johnson66}.

\begin{deluxetable}{lrrrrrrrrrrrrllr}
\tablewidth{0pt}
\tablecaption{\label{table.supergiants} Infrared colors and \BCK\ of supergiant stars.}
\tablehead{
\colhead{Sp.}& 
\colhead{T$_{eff}$}& 
\colhead{BC$_K$}& 
\colhead{$J-K$}& 
\colhead{$H-K$}& 
\colhead{Reference}& 
}
\startdata
O3 &  42551  &  $-$4.69  &    $-$0.21  &      $-$0.10 &\citet{martins05},\citet{martins06}\\ 
O4 &  40702  &  $-$4.55  &    $-$0.21  &      $-$0.10 &\citet{martins05},\citet{martins06}\\ 
O5 &  38520  &  $-$4.40  &    $-$0.21  &      $-$0.10 &\citet{martins05},\citet{martins06}\\ 
O6 &  35747  &  $-$4.25  &    $-$0.21  &      $-$0.10 &\citet{martins05},\citet{martins06}\\ 
O7 &  33326  &  $-$4.09  &    $-$0.21  &      $-$0.10 &\citet{martins05},\citet{martins06}\\ 
O8 &  31009  &  $-$3.93  &    $-$0.21  &      $-$0.10 &\citet{martins05},\citet{martins06}\\ 
O9 &  29569  &  $-$3.75  &    $-$0.21  &      $-$0.10 &\citet{martins05},\citet{martins06}\\ 
B0 &  27500  &  $-$3.30  &    $-$0.11  &      $-$0.04 &\citet{crowther06}, \citet{bibby08}\\ 
B1 &  21500  &  $-$2.65  &    $-$0.09  &      $-$0.03 &\citet{crowther06}, \citet{bibby08}\\ 
B2 &  18500  &  $-$2.10  &    $-$0.07  &      $-$0.03 &\citet{crowther06}, \citet{bibby08}\\ 
B3 &  15500  &  $-$1.70  &    $-$0.04  &      $-$0.03 &\citet{crowther06}, \citet{bibby08}\\ 
B5 &  13700  &  $-$0.95  &       0.00  &      $-$0.01 &\citet{humphreys84},\citep{koornneef83}\\
B8 &  10900  &  $-$0.47  &       0.05  &	 0.00 &\citet{humphreys84},\citep{koornneef83}\\
B9 &  10250  &  $-$0.27  &       0.07  &	 0.01 &\citet{humphreys84},\citep{koornneef83}\\
A0 &   9500  &  $-$0.09  &       0.09  &	 0.01 &\citet{humphreys84},\citep{koornneef83}\\
A2 &   9100  &     0.14  &       0.11  &	 0.02 &\citet{humphreys84},\citep{koornneef83}\\
A5 &   8500  &     0.36  &       0.12  &	 0.02 &\citet{humphreys84},\citep{koornneef83}\\
\enddata		 																         
\end{deluxetable}

\begin{deluxetable}{lrrrrrrrrrrrrllr}
\tablewidth{0pt}
\tablecaption{\label{table.giants} Infrared colors and \BCK\ of giant stars.}
\tablehead{
\colhead{Sp.}& 
\colhead{T$_{eff}$}& 
\colhead{BC$_K$}& 
\colhead{$J-K$}& 
\colhead{$H-K$}& 
\colhead{Reference}& 
}
\startdata
O3 &  42942&   $-4.85$ &  $-0.21$  &  $-0.10$&\citet{martins05},\citet{martins06}\\ 
O4 &  41486&   $-4.70$ &  $-0.21$  &  $-0.10$&\citet{martins05},\citet{martins06}\\ 
O5 &  39507&   $-4.54$ &  $-0.21$  &  $-0.10$&\citet{martins05},\citet{martins06}\\ 
O6 &  36673&   $-4.37$ &  $-0.21$  &  $-0.10$&\citet{martins05},\citet{martins06}\\ 
O7 &  34638&   $-4.19$ &  $-0.21$  &  $-0.10$&\citet{martins05},\citet{martins06}\\ 
O8 &  32573&   $-4.00$ &  $-0.21$  &  $-0.10$&\citet{martins05},\citet{martins06}\\ 
O9 &  30737&   $-3.80$ &  $-0.21$  &  $-0.10$&\citet{martins05},\citet{martins06}\\ 
B0 &  30300&   $-3.59$ &  $-0.18$  &  $-0.09$&\citet{humphreys84},\citep{wegner94}\\ 
B1 &  21100&   $-2.65$ &  $-0.19$  &  $-0.10$&\citet{humphreys84},\citep{wegner94}\\ 
B2 &  18000&   $-2.16$ &  $-0.18$  &  $-0.10$&\citet{humphreys84},\citep{wegner94}\\
B3 &  17100&   $-1.96$ &  $-0.12$  &  $-0.06$&\citet{humphreys84},\citep{wegner94}\\ 
B5 &  16300&   $-1.81$ &  $-0.10$  &  $-0.06$&\citet{humphreys84},\citep{wegner94}\\
B8 &  12550&   $-0.94$ &  $-0.03$  &  $-0.02$&\citet{humphreys84},\citep{wegner94}\\
B9 &  11400&   $-0.55$ &  $-0.02$  &  $-0.01$&\citet{humphreys84},\citep{wegner94}\\
\enddata		 																         
\end{deluxetable}

\begin{deluxetable}{lrrrrrrrrrrrrllr}
\tablewidth{0pt}
\tablecaption{\label{table.dwarfs} Infrared colors and \BCK\ of dwarf stars.}
\tablehead{
\colhead{Sp.}& 
\colhead{T$_{eff}$}& 
\colhead{BC$_K$}& 
\colhead{$J-K$}& 
\colhead{$H-K$}& 
\colhead{Reference}& 
}
\startdata
O3 &     44616 &      $-4.76$  &  $-0.19$  &    $-0.04$ &\citet{martins05},\citet{lejeune01}\\ 
O4 &     43419 &      $-4.70$  &  $-0.19$  &    $-0.04$ &\citet{martins05},\citet{lejeune01}\\ 
O5 &     41540 &      $-4.62$  &  $-0.18$  &    $-0.04$ &\citet{martins05},\citet{lejeune01}\\ 
O6 &     38151 &      $-4.39$  &  $-0.17$  &    $-0.04$ &\citet{martins05},\citet{lejeune01}\\ 
O7 &     35531 &      $-4.15$  &  $-0.17$  &    $-0.04$ &\citet{martins05},\citet{lejeune01}\\ 
O8 &     33383 &      $-3.98$  &  $-0.16$  &    $-0.04$ &\citet{martins05},\citet{lejeune01}\\ 
O9 &     31524 &      $-3.86$  &  $-0.15$  &    $-0.04$ &\citet{martins05},\citet{lejeune01}\\ 
B0 &     29600 &      $-3.72$  &  $-0.15$  &    $-0.04$ &\citet{humphreys84},\citet{lejeune01}\\
B1 &     24150 &      $-3.21$  &  $-0.13$  &    $-0.03$ &\citet{humphreys84},\citet{lejeune01}\\
B2 &     19700 &      $-2.60$  &  $-0.12$  &    $-0.03$ &\citet{humphreys84},\citet{lejeune01}\\
B3 &     18700 &      $-2.46$  &  $-0.11$  &    $-0.02$ &\citet{humphreys84},\citet{lejeune01}\\
B5 &     13800 &      $-1.43$  &  $-0.06$  &    $-0.01$ &\citet{johnson66},\citet{lejeune01}\\
B8 &     12200 &      $-1.00$  &  $-0.03$  &    $-0.00$ &\citet{johnson66},\citet{lejeune01}\\
B9 &     10600 &      $-0.52$  &  $-0.00$  &    $ 0.00$ &\citet{johnson66},\citet{lejeune01}\\
A0 &      9850 &      $-0.25$  &  $ 0.02$  &    $ 0.01$ &\citet{johnson66},\citet{lejeune01}\\
A2 &      9120 &      $-0.03$  &  $ 0.04$  &    $ 0.02$ &\citet{johnson66},\citet{lejeune01}\\
A5 &      8260 &      $ 0.28$  &  $ 0.08$  &    $ 0.03$ &\citet{johnson66},\citet{lejeune01}\\
\enddata		 																         
\end{deluxetable}

\begin{deluxetable}{lrrrrrrrrrrrrllr}
\tablewidth{0pt}
\tablecaption{\label{table.othertype} Infrared colors and \BCK\ adopted for RSGs and WRs.}
\tablehead{
\colhead{Sp.}& 
\colhead{T$_{eff}$}& 
\colhead{BC$_K$}& 
\colhead{$J-K$}& 
\colhead{$H-K$}& 
\colhead{Reference}& 
}
\startdata
K2I   &     4015  &      $+2.5 $  &  $+0.65$  &    $+0.13$ &\citet{levesque05},\citet{koornneef83}\\ 
M1I   &     3745  &      $+2.7 $  &  $+1.00$  &    $+0.22$ &\citet{levesque05},\citet{koornneef83}\\ 
M2.5I &     3615  &      $+2.8 $  &  $+1.06$  &    $+0.27$ &\citet{levesque05},\citet{koornneef83}\\ 
M3.5I &     3550  &      $+2.9 $  &  $+1.16$  &    $+0.28$ &\citet{levesque05},\citet{koornneef83}\\ 
WN7o  &           &      $-3.9 $  &  $+0.13$  &    $+0.11$ &\citet{crowther06}\\
WN7b  &           &      $-3.5 $  &  $+0.37$  &    $+0.27$ &\citet{crowther06}\\
\enddata		 																         
\end{deluxetable}
           

\end{document}